\documentclass[british,aps, onecolumn,notitlepage, pre,superscriptaddress, notitlepage, longbibliography , dvipsnames]{revtex4-2}
\usepackage[T1]{fontenc}
\usepackage{array}
\usepackage{booktabs}
\usepackage{bm}
\usepackage{amsmath}
\usepackage{amssymb}

\makeatletter
\newcommand{\fmarki}{\ensuremath{\dagger}}
                
\def\@fnsymbol#1{{\ifcase#1\or \fmarki\or \fmarkii\or \fmarkiii\or \fmarkiv\or \fmarkv\or \fmarkvi\or \fmarkvii\or \fmarkviii\or \fmarkix \else\@ctrerr\fi}}
\makeatother

\providecommand{\tabularnewline}{\\}

\usepackage{amsmath}
\usepackage{subfigure}
\usepackage{graphicx, mathtools}
\usepackage[normalem]{ulem}
\usepackage[dvipsnames]{xcolor}
\usepackage[colorlinks=true,linkcolor=MidnightBlue,urlcolor=MidnightBlue,citecolor=MidnightBlue,anchorcolor=MidnightBlue]{hyperref}
\usepackage[T1]{fontenc}
\usepackage[utf8]{inputenc}
\makeatother

\usepackage{babel}
\begin{document}
\title{Fluctuating hydrodynamics of an autophoretic particle near a permeable
interface}
\author{G\"{u}nther Turk}
\email{guenther.turk@princeton.edu}

\affiliation{Princeton Materials Institute, Princeton University, Princeton, NJ,
08544, United States of America}
\author{Ronojoy Adhikari}
\affiliation{DAMTP, Centre for Mathematical Sciences, University of Cambridge,
Wilberforce Road, Cambridge CB3 0WA, United Kingdom}
\affiliation{\textsuperscript{}The Institute of Mathematical Sciences-HBNI, CIT
Campus, Chennai 600113, India }
\author{Rajesh Singh}
\affiliation{Department of Physics, IIT Madras, Chennai 600036, India}
\begin{abstract}
We study the autophoretic motion of a spherical active particle interacting
chemically and hydrodynamically with its fluctuating environment in
the limit of rapid diffusion and slow viscous flow. Then, the chemical
and hydrodynamic fields can be expressed in terms of integrals. The
resulting boundary-domain integral equations provide a direct way
of obtaining the traction on the particle, requiring the solution
of linear integral equations. An exact solution for the chemical and
hydrodynamic problems is obtained for a particle in an unbounded domain.
For motion near boundaries, we provide corrections to the unbounded
solutions in terms of chemical and hydrodynamic Green's functions,
preserving the dissipative nature of autophoresis in a viscous fluid
for all physical configurations. Using this, we give the fully stochastic
update equations for the Brownian trajectory of an autophoretic particle
in a complex environment. First, we analyse the Brownian dynamics
of particles capable of complex motion in the bulk. We then introduce
a chemically permeable planar surface of two immiscible liquids in
the vicinity of the particle and provide explicit solutions to the
chemo-hydrodynamics of this system. Finally, we study the case of
an isotropically phoretic particle hovering above an interface as
a function of interfacial solute permeability and viscosity contrast.\\
\\
DOI: \href{https://doi.org/10.1017/jfm.2024.661}{10.1017/jfm.2024.661}
\end{abstract}
\maketitle

\section{Introduction}

Autophoretic motion comprises the propulsion of particles due to self-generated
gradients \citep{andersonColloidTransportInterfacial1989,paxton2006chemical,moranPhoreticSelfPropulsion2017,ebbens2010pursuit},
typically on an energy scale comparable to that of thermal fluctuations
\citep{batchelor1976developments,grahamMicrohydrodynamicsBrownianMotion2018}.
This self-propulsion mechanism allows systems of phoretic particles
to mimic the locomotion of microorganisms \citep{goldstein2015green,brennen1977},
making them useful in the study of the fundamental principles of motility
and collective behaviour \citep{palacci2013living,illien2017fuelled,shaebaniComputationalModelsActive2020,kumarEmergentDynamicsDue2024,zottlModelingActiveColloids2023}.
In particular, the study of interactions of autophoretic particles
with nearby boundaries is relevant in micro-fluidics, biophysics and
surface science \citep{kreuterTransportPhenomenaDynamics2013,uspalSelfpropulsionCatalyticallyActive2015,ibrahimDynamicsSelfphoreticJanus2015,shenHydrodynamicInteractionSelfpropelling2018,singhCompetingChemicalHydrodynamic2019,thutupalli2018FIPS}.

Our goal is thus to formulate an effective description in which Brownian
motion and autophoresis of active particles can be studied when suspended
in a complex environment. Models for self-diffusiophoresis typically
assume that chemical gradients generated by the particle induce an
osmotic pressure, which is balanced by viscous stresses driving an
effective slip flow confined to a thin layer
at the surface of the particle \citep{andersonMotionParticleGenerated1982}.
This sets the surrounding fluid in motion, with fluid stresses reacting
back on the particle and setting it in motion. To compute the particle
dynamics usually requires solving for the concentration field and
the fluid flow in the bulk, and subsequently obtaining the stresses
on the particle by matching all relevant boundary conditions \citep{golestanian2005propulsion,golestanian2007designing}.
Instead, by using a boundary-domain integral approach, we directly
obtain the concentration distribution and the resulting traction (force
per unit area) on the surface of the particle, obviating the need
for solving the governing equations in the bulk. Compared with more
conventional kinematic approaches \citep{lighthillSquirmingMotionNearly1952,pakGeneralizedSquirmingMotion2014},
it is then straightforward to incorporate thermal fluctuations in
the surrounding fluid as Brownian stresses on the particle. The latter
have been studied extensively for suspensions of colloidal particles
\citep{einstein1905theory,zwanzig1964hydrodynamic,chowSimultaneousTranslationalRotational1973,hinchApplicationLangevinEquation1975,ermakBrownianDynamicsHydrodynamic1978,ladd1994numericala,cichockiFrictionMobilityColloidal2000,keavenyFluctuatingForcecouplingMethod2014,delmotteSimulatingBrownianSuspensions2015,singhFluctuatingHydrodynamicsBrownian2017,elfringActiveStokesianDynamics2022,turkStokesTractionActive2022,mozaffariSelfpropelledColloidalParticle2018,baoFluctuatingBoundaryIntegral2018,westwoodGeneralisedDriftcorrectingTime2022},
highlighting that any acceptable approximation of the colloidal diffusion
matrix in Brownian dynamics modelling must remain positive-definite
for all physical configurations \citep{wajnrybBrownianDynamicsDivergence2004,wajnrybGeneralizationRotnePrager2013}.
Based on a Galerkin-Jacobi iterative method, the analytical expressions
we provide naturally satisfy this condition.

The fields generated by and the resulting stresses on autophoretic
particles are well known in an unbounded fluid \citep{golestanian2007designing,ebbens2010pursuit,illien2017fuelled,lisickiAutophoreticMotionThree2018},
or when confined either by no-slip walls that are impermeable to the
solutes \citep{crowdyWallEffectsSelfdiffusiophoretic2013,ibrahimDynamicsSelfphoreticJanus2015,uspalSelfpropulsionCatalyticallyActive2015,ibrahimHowWallsAffect2016,mozaffariSelfdiffusiophoreticColloidalPropulsion2016,daddi-moussa-iderSwimmingTrajectoriesThreesphere2018,kanso2019phoretic,singhCompetingChemicalHydrodynamic2019},
or by chemically patterned boundaries \citep{uspalActiveJanusColloids2019}.
In this paper we formulate a general framework for finding the full
chemo-hydrodynamics of a particle in an arbitrary complex environment
in terms of chemical and hydrodynamic Green's functions (the fields
generated by Dirac delta function sources \citep{ladyzhenskaya1969}).
Using this, we provide analytically the dynamics of a phoretic particle
in the proximity of a chemically permeable liquid-liquid interface
separating the suspending domain from a second, immiscible liquid
phase. Assuming a large capillary number, we restrict our considerations
to a planar interface. This is particularly relevant for studies on
particle aggregation near fluid-fluid interfaces and free surfaces
\citep{chen2015dynamic,hokmabadSpontaneouslyRotatingClusters2022},
with a permeable interface being a plausible model of biofilms and
hydrogels \citep{wichterleHydrophilicGelsBiological1960,berke2008hydrodynamic}. 

The rest of the paper is organised as follows. In Section \ref{sec:chemo-hydrodynamics},
we review the chemo-hydrodynamic problem of autophoresis in a fluctuating
environment and its formal solution via the boundary-domain integral
representation of Laplace and Stokes equations. In Section \ref{sec:solutionYL},
we then use a Galerkin discretisation to project the formal solution
onto a basis of tensor spherical harmonics (TSH), finding an exact
and an approximate solution to the full chemo-hydrodynamic problem
far away from and near boundaries, respectively. We provide the stochastic
update equations for thermally agitated autophoresis in complex environments.
In Section \ref{sec:Applications}, we apply these equations to the
study of three representative examples. First, we consider systematically
patterned particle surfaces, which we confirm can lead to complex
phoretic motion even in the absence of boundaries \citep{lisickiAutophoreticMotionThree2018}.
We study the effect of thermal fluctuations on the resulting particle
motion in a bulk fluid. In the vicinity of the particle we then introduce
the presence of a plane surface of two immiscible liquids that is
permeable to the solutes. For this system, we obtain explicit forms
of the relevant chemical and hydrodynamic connectors. We demonstrate
our analytical results by numerically investigating the chemo-hydrodynamic
effects the interface has on the dynamics of a nearby autophoretic
particle. This includes an analysis of the hovering state of a phoretic
particle above an interface as a function of particle activity, and
interfacial properties. We conclude with a brief discussion of our
results and potential future applications thereof in Section \ref{sec:Summary-and-outlook}.

\section{chemo-hydrodynamics\label{sec:chemo-hydrodynamics}}

\begin{table}
\begin{tabular}{>{\centering}m{7cm}>{\centering}m{7cm}}
\toprule 
\textbf{Chemical problem} & \textbf{Hydrodynamic problem}\tabularnewline
\midrule
\midrule 
\addlinespace
$\qquad\boldsymbol{\nabla}\cdot\boldsymbol{j}=0$\hfill{}(Ia) & $\qquad\boldsymbol{\nabla}\cdot\boldsymbol{\sigma}+\boldsymbol{\xi}=0$\hfill{}(Id)\tabularnewline\addlinespace
\addlinespace
$\qquad\boldsymbol{j}=-D\boldsymbol{\nabla}c$\hfill{}(Ib) & $\qquad\boldsymbol{\sigma}=-p\boldsymbol{I}+\eta\left(\boldsymbol{\nabla}\boldsymbol{v}+\left(\boldsymbol{\nabla}\boldsymbol{v}\right)^{\text{tr}}\right)$\hfill{}(Ie)\tabularnewline\addlinespace
\midrule 
\addlinespace
$\qquad\boldsymbol{j}(\boldsymbol{R}+\boldsymbol{b})\cdot\hat{\boldsymbol{b}}=j^{\mathcal{A}}(\boldsymbol{b})\qquad$(specified)\hfill{}(Ic) & $\qquad\boldsymbol{v}(\boldsymbol{R}+\boldsymbol{b})=\boldsymbol{v}^{\mathcal{D}}(\boldsymbol{b})+\boldsymbol{v}^{\mathcal{A}}(\boldsymbol{b})$\hfill{}(If)\tabularnewline\addlinespace
\midrule 
\addlinespace
\multicolumn{2}{c}{\textbf{\hspace{1cm}Chemo-hydrodynamic coupling}:\textbf{$\quad$}
$\boldsymbol{v}^{\mathcal{A}}(\boldsymbol{b})=\boldsymbol{\chi}\left[c\right],$$\qquad\boldsymbol{\chi}=\mu_{c}(\boldsymbol{b})\big(\boldsymbol{I}-\hat{\boldsymbol{b}}\hat{\boldsymbol{b}}\big)\cdot\boldsymbol{\nabla}$\hfill{}(Ig)}\tabularnewline\addlinespace
\bottomrule
\end{tabular}\caption{\label{tab:CH-diff}\textbf{Governing differential laws.} This table
summarises the chemo-hydrodynamic coupling at the surface of an autophoretic
particle, an example of two three-dimensional partial differential
equations, namely Laplace equation for the concentration field (Ia)
and Stokes equation for the fluid flow and pressure in a fluid with
thermal force density $\boldsymbol{\xi}$ (Id), coupling on a two-dimensional
surface only \citep{melcher1969electrohydrodynamics}. The chemo-hydrodynamic
coupling (Ig) leads to the specified active flux $j^{\mathcal{A}}$
driving a slip flow $\boldsymbol{v}^{\mathcal{A}}$ in a thin layer
at the surface of the particle with specified phoretic mobility $\mu_{c}$,
finally driving the fluid surrounding the particle and causing self-propulsion.
A passive particle is a rigid sphere of radius $b$ with the boundary
condition: $\boldsymbol{v}^{\mathcal{D}}(\boldsymbol{b})=\boldsymbol{V}+\boldsymbol{\Omega}\times\boldsymbol{b}$,
where $\boldsymbol{V}$ is the velocity and $\boldsymbol{\Omega}$
is the angular velocity of the particle. An active particle is modelled
as a sphere with boundary condition (If), which comprises both slip
$\boldsymbol{v}^{\mathcal{A}}(\boldsymbol{b})$ and rigid-body motion
$\boldsymbol{v}^{\mathcal{D}}(\boldsymbol{b})$ \citep{andersonColloidTransportInterfacial1989,ebbens2010pursuit}.}
\end{table}
\begin{table}
\begin{tabular}{>{\centering}m{8cm}>{\centering}m{8cm}}
\toprule 
\textbf{Chemical problem} & \textbf{Hydrodynamic problem}\tabularnewline
\midrule
\midrule 
\addlinespace
$\qquad\tfrac{1}{2}c=c^{\infty}+\mathcal{H}\left[j^{\mathcal{A}}\right]+\mathcal{L}\left[c\right]$\hfill{}(IIa)\smallskip{}
 & $\qquad\tfrac{1}{2}\boldsymbol{v}=-\boldsymbol{\mathcal{G}}\left[\boldsymbol{f}\right]+\boldsymbol{\mathcal{K}}\left[\boldsymbol{v}\right]+\boldsymbol{u}\left[\boldsymbol{\xi}\right],$\hfill{}(IIf)\smallskip{}
\tabularnewline\addlinespace
\addlinespace
$\qquad\mathcal{H}\left[j^{\mathcal{A}}\right]=\int H(\boldsymbol{r},\tilde{\boldsymbol{r}})j^{\mathcal{A}}(\tilde{\boldsymbol{r}})\mathrm{d}S'$\hfill{}(IIb) & $\qquad\boldsymbol{\mathcal{G}}\left[\boldsymbol{f}\right]=\int\boldsymbol{G}(\boldsymbol{r},\tilde{\boldsymbol{r}})\cdot\boldsymbol{f}(\tilde{\boldsymbol{r}})\mathrm{d}S'$\hfill{}(IIg)\tabularnewline\addlinespace
\addlinespace
$\qquad\mathcal{L}\left[c\right]=\int c(\tilde{\boldsymbol{r}})\boldsymbol{L}(\boldsymbol{r},\tilde{\boldsymbol{r}})\cdot\hat{\boldsymbol{b}}'\mathrm{d}S'$\hfill{}(IIc) & $\qquad\boldsymbol{\mathcal{K}}\left[\boldsymbol{v}\right]=\int\boldsymbol{v}(\tilde{\boldsymbol{r}})\cdot\boldsymbol{K}(\tilde{\boldsymbol{r}},\boldsymbol{r})\cdot\hat{\boldsymbol{b}}'\mathrm{d}S'$\hfill{}(IIh)\tabularnewline\addlinespace
\addlinespace
 & $\qquad\boldsymbol{u}\left[\boldsymbol{\xi}\right]=\int\boldsymbol{G}(\mathbf{r},\tilde{\mathbf{r}})\cdot\boldsymbol{\xi}(\tilde{\mathbf{r}})\,{\rm d}V',$\hfill{}(IIi)\tabularnewline\addlinespace
\midrule 
\addlinespace
$\qquad c=\left(\tfrac{1}{2}-\mathcal{L}\right)^{-1}\left\{ \left[c^{\infty}\right]+\mathcal{H}\left[j^{\mathcal{A}}\right]\right\} $\hfill{}(IId) & $\qquad\boldsymbol{f}=-\boldsymbol{\mathcal{G}}^{-1}\left\{ \left[\boldsymbol{v}^{\mathcal{D}}\right]+\left(\tfrac{1}{2}\boldsymbol{I}-\boldsymbol{\mathcal{K}}\right)\left[\boldsymbol{v}^{\mathcal{A}}\right]-\boldsymbol{u}\left[\boldsymbol{\xi}\right]\right\} $\hfill{}(IIj)\tabularnewline\addlinespace
\addlinespace
$\qquad\,\,\,\,\equiv\zeta\left[c^{\infty}\right]+\mathcal{E}\left[j^{\mathcal{A}}\right]$\hfill{}(IIe) & $\qquad\,\,\,\,\,\equiv-\boldsymbol{\gamma}\left[\boldsymbol{v}^{\mathcal{D}}\right]-\hat{\boldsymbol{\gamma}}\left[\boldsymbol{v}^{\mathcal{A}}\right]+\boldsymbol{\gamma}\left[\boldsymbol{u}\left[\boldsymbol{\xi}\right]\right]$\hfill{}(IIk)\tabularnewline\addlinespace
\midrule 
\addlinespace
\multicolumn{2}{c}{\hspace{3cm}\textbf{Chemo-hydrodynamic coupling:$\quad$} $\boldsymbol{v}^{\mathcal{A}}(\boldsymbol{b})=\boldsymbol{\chi}\left\{ \zeta\left[c^{\infty}\right]+\mathcal{E}\left[j^{\mathcal{A}}\right]\right\} $\hfill{}(IIl)}\tabularnewline\addlinespace
\bottomrule
\end{tabular}

\caption{\label{tab:CH-int}\textbf{Governing integral laws. }This table summaries
the formal solutions to the boundary-domain integral equations corresponding
to Laplace equation for the concentration $c$ (IIa) and Stokes equation
for the traction (force per unit area) $\boldsymbol{f}$ (IIf) on
the surface of the particle. Here $\boldsymbol{r},\tilde{\boldsymbol{r}}\in S$,
where $\boldsymbol{r}=\boldsymbol{R}+\boldsymbol{b}$ and $\tilde{\boldsymbol{r}}=\boldsymbol{R}+\boldsymbol{b}'$
are the field and source points at the surface $S$ of the particle
centred at $\boldsymbol{R}$, respectively, and $\int\mathrm{d}S'$
implies an integration over $\tilde{\boldsymbol{r}}$. In Eqs. (IId)
and (IIj) we give the solutions for the concentration and the traction
in terms of integral linear operators. We have used the fact that
rigid-body motion $\boldsymbol{v}^{\mathcal{D}}$ lies in the eigenspectrum
of the double-layer operator (IIh) with an eigenvalue $-1/2$ \citep{kimEllipsoidalMicrohydrodynamicsElliptic2015}.
In the formal solutions we have introduced operators representing
the linear response to a background concentration field $\zeta$,
the so-called elastance $\mathcal{E}$, the rigid-body friction $\boldsymbol{\gamma}$,
and the friction due to surface slip $\hat{\boldsymbol{\gamma}}$.
Inserting the operator solution for the concentration in Eq. (IIe)
into Eq. (Ig) in Table \ref{tab:CH-diff} for the chemo-hydrodynamic
coupling at the surface of an autophoretic particle, we find Eq. (IIl)
for the surface slip $\boldsymbol{v}^{\mathcal{A}}(\boldsymbol{b})$.}
\end{table}
We consider a spherical autophoretic particle of radius $b$, suspended
in an incompressible fluid ($\boldsymbol{\nabla}\cdot\boldsymbol{v}=0$,
where $\boldsymbol{v}$ is the flow field) of viscosity $\eta$ at
low Reynolds number. Thermal fluctuations of the fluid at equilibrium
are modelled by a zero-mean Gaussian random field $\boldsymbol{\xi}$,
the thermal force acting on the particle, whose variance is given
by a fluctuation-dissipation relation \citep{hauge1973fluctuating,fox1970contributions,bedeaux1974brownian,roux1992brownian,zwanzig1964hydrodynamic}.
In Table \ref{tab:CH-diff}, we summarise the differential laws governing
the chemo-hydrodynamics of this system. We denote fields defined on
the surface of spherical particles as functions of the radius vector
$\boldsymbol{b}$ of the sphere, where $\hat{\boldsymbol{b}}=\boldsymbol{b}/b$
is the unit outward normal to the surface, pointing into the fluid
and with $b=|\boldsymbol{b}$|. We assume a negligibly small P\'{e}clet
number, thus ignoring distortions induced by the flow on the solute
concentration \citep{michelinSpontaneousAutophoreticMotion2013,morozovNonlinearDynamicsChemicallyactive2019}.
Additionally, we assume that solute diffusion takes place on much
shorter time scales than Brownian motion of the autophoretic particle,
which in turn takes place on much shorter time scales than its rigid-body motion. The chemical problem is then represented by the Laplace
equation for the concentration field $c$, for ideal solutions equivalent
to a divergence-free chemical flux $\boldsymbol{j}$ in Eq. (Ia),
where $D$ is the solute diffusivity in the fluid. In Eq. (Ic) the
normal component of the flux at the surface of the particle $j^{\mathcal{A}}(\boldsymbol{b})$
is specified.

Surface gradients of the generated concentration field induce a mass
transport of solute, thus driving a fluid flow confined to a thin layer
at the surface of the particle. This is modelled by a slip $\boldsymbol{v}^{\mathcal{A}}$
in the chemo-hydrodynamic coupling in Eq. (Ig). Here, $\mu_{c}$ is
the particle-specific phoretic mobility, which incorporates the solute-colloid
interactions. We assume that the solute is uncharged (neutral diffusiophoresis)
\citep{prieveMotionParticleGenerated1984,velegolOriginsConcentrationGradients2016,yangAutophoresisTwoAdsorbing2019}.
The slip is incorporated in the velocity boundary condition in Eq.
(If), alongside rigid-body motion $\boldsymbol{v}^{\mathcal{D}}$
of the particle. Finally, the particle sets the surrounding fluid
in motion (via the slip or rigid-body motion due to external forces
and torques), hydrodynamically interacting with its surroundings via
the Stokes equation (Id). Therein, we have defined the Cauchy stress
tensor $\boldsymbol{\sigma}$, containing contributions from the isotropic
fluid pressure $p$ and from spatial variations in the flow field.
Here $\boldsymbol{I}$ is the identity tensor.

In Table \ref{tab:CH-int}, we summarise the boundary-domain integral
equations (BIEs) corresponding to the Laplace and Stokes equations,
and their formal solution in terms of integral linear operators. The
BIE (IIa) for the concentration at the surface of the particle is
given in terms of a background concentration field $c^{\infty}(\boldsymbol{r})$,
the single-layer operator $\mathcal{H}\left[j^{\mathcal{A}}\right]$
 and the double-layer operator $\mathcal{L}\left[c\right]$. This naming
convention of the integral operators is by analogy with potential
theory \citep{jackson1962classical,kim2005}. The integral kernels
contain the concentration Green's function $H$ and its gradient $\boldsymbol{L}$.
Due to linearity of the Laplace equation, we can find the solution in
Eq. (IId) for the concentration, containing the operator $\zeta$
for the linear response to a background chemical field and the so-called
elastance operator $\mathcal{E}$. The naming convention of the latter
originates from Maxwell, who in his study of the capacitance of a
system of spherical conductors coined the term elastance for the isotropic
part of the tensor $\boldsymbol{\mathcal{E}}$ \citep{maxwell1881treatise}.

The corresponding BIE of fluctuating Stokes flow (IIf) is a sum of
the single-layer operator $\boldsymbol{\mathcal{G}}\left[\boldsymbol{f}\right]$
acting on the surface traction (force per unit area) on the particle,
given by $\boldsymbol{f}=\boldsymbol{\sigma}\cdot\hat{\boldsymbol{b}}$,
the double-layer operator $\boldsymbol{\mathcal{K}}\left[\boldsymbol{v}\right]$
\citep{lorentz1896eene,fkg1930bandwertaufgaben,ladyzhenskaya1969,youngren1975stokes,zick1982stokes,pozrikidisBoundaryIntegralSingularity1992,muldowney1995spectral,cheng2005heritage,leal2007advanced,singhManybodyMicrohydrodynamicsColloidal2015} 
and the Brownian velocity field $\boldsymbol{u}\left[\boldsymbol{\xi}\right]$
\citep{singhFluctuatingHydrodynamicsBrownian2017}. The integral kernels
contain the Green's function $\boldsymbol{G}$ of the Stokes equation
and the stress tensor $\boldsymbol{K}$ associated with it. Linearity
of the Stokes equation allows us to formally solve the BIE, introducing
the friction operators $\boldsymbol{\gamma}$ and $\hat{\boldsymbol{\gamma}}$
due rigid-body motion and slip, respectively. They can be distinguished
by a non-trivial contribution of the double-layer integral to the
latter \citep{turkStokesTractionActive2022}. Finally, the solutions
to the chemical and hydrodynamic problems are coupled via the boundary
condition (IIl).

In the following, an autophoretic particle is fully specified by its
surface flux $j^{\mathcal{A}}$ and phoretic mobility $\mu_{c}$, as
indicated in Table \ref{tab:CH-diff}. Our aim is to find its dynamics,
governed by Newton's laws
\begin{equation}
m\dot{\boldsymbol{V}}=\boldsymbol{F}^{H}+\boldsymbol{F}^{P}+\hat{\boldsymbol{F}},\qquad I\dot{\boldsymbol{\Omega}}=\boldsymbol{T}^{H}+\boldsymbol{T}^{P}+\hat{\boldsymbol{T}}.\label{eq:Newton}
\end{equation}
Here, $m$ and $I$ are the particle mass and moment of inertia, respectively,
and a dotted variable implies a time derivative. Body forces and torques
are denoted by $\boldsymbol{F}^{P}$ and $\boldsymbol{T}^{P}$, and
the hydrodynamic and fluctuating contributions are defined in terms
of the traction on the particle
\begin{equation}
\boldsymbol{F}^{H}=\int\boldsymbol{f}^{H}{\rm d}S,\qquad\boldsymbol{T}^{H}=\int\boldsymbol{b}\times\boldsymbol{f}^{H}{\rm d}S,\qquad\qquad\hat{\boldsymbol{F}}=\int\hat{\boldsymbol{f}}{\rm d}S,\qquad\hat{\boldsymbol{T}}=\int\boldsymbol{b}\times\hat{\boldsymbol{f}}{\rm d}S,\label{eq:forcetorque}
\end{equation}
where the total surface traction on the particle is the sum $\boldsymbol{f}=\boldsymbol{f}^{H}+\hat{\boldsymbol{f}}$.
We define the hydrodynamic traction due to rigid-body and active interactions
as $\boldsymbol{f}^{H}$ and the Brownian traction due to thermal
fluctuations in the fluid as $\hat{\bm{f}}$ such that
\begin{eqnarray}
\boldsymbol{f}^{H} & = & -\boldsymbol{\gamma}\left[\boldsymbol{v}^{\mathcal{D}}\right]-\hat{\boldsymbol{\gamma}}\left[\boldsymbol{v}^{\mathcal{A}}\right],\qquad\hat{\boldsymbol{f}}=\boldsymbol{\gamma}\left[\boldsymbol{u}\left[\boldsymbol{\xi}\right]\right].
\end{eqnarray}
It is known that the latter are zero-mean random variables with variances
fixed by a fluctuation-dissipation relation \citep{zwanzig1964hydrodynamic,chowSimultaneousTranslationalRotational1973}.
By linearity of the governing equations, the hydrodynamic and Brownian
contributions can be solved for independently and the fluid degrees
of freedom can be eliminated exactly, yielding the Brownian dynamics
of the active particle. 

\section{Solution in an irreducible basis\label{sec:solutionYL}}

In this section, we write the formal solutions to Laplace and Stokes
equation in Table \ref{tab:CH-int}, Eqs. (IIe) and (IIk), in an irreducible
basis, thus transforming the integral operator equations into linear
systems, for which we give explicit solutions. We choose a basis of
TSH, defined by
\[
Y_{\alpha_{1}\dots\alpha_{l}}^{(l)}(\hat{\boldsymbol{b}})=(2l-1)!!\Delta_{\alpha_{1}\dots\alpha_{l},\beta_{1}\dots\beta_{l}}^{(l)}\hat{b}_{\beta_{1}}\dots\hat{b}_{\beta_{l}}=(-1)^{l}\,b^{l+1}\,\nabla_{\alpha_{1}}\dots\nabla_{\alpha_{l}}\frac{1}{b},
\]
where $\boldsymbol{\Delta}{}^{(l)}$ is a rank-$2l$ tensor, which
projects a tensor of rank-$l$ onto its symmetric and traceless part
\citep{hess2015tensors}.

\subsection{Chemical problem\label{subsec:Boundary-integral-solution}}

To project Eq. (IIe) for the concentration at the surface of the particle
onto a linear system, we expand the boundary fields

\begin{gather}
c(\boldsymbol{b})=\sum_{q=0}^{\infty}w_{q}\boldsymbol{C}^{(q)}\odot\boldsymbol{Y}^{(q)}(\hat{\boldsymbol{b}}),\qquad j^{\mathcal{A}}(\boldsymbol{b})=\sum_{q=0}^{\infty}\tilde{w}_{q}\boldsymbol{J}^{(q)}\odot\boldsymbol{Y}^{(q)}(\hat{\boldsymbol{b}}).\label{eq:exp-c}
\end{gather}
The product denoted by $\odot$ implies a maximal contraction of Cartesian
indices (a \textbf{$q$}-fold contraction between a tensor of rank-$q$
and another one of higher rank) and we have defined
\begin{equation}
w_{q}=\frac{1}{q!\left(2q-1\right)!!},\quad\tilde{w}_{q}=\frac{2q+1}{4\pi b^{2}}.
\end{equation}
The expansion coefficients $\boldsymbol{C}^{(q)}$ and $\boldsymbol{J}^{(q)}$
are symmetric and traceless tensors of rank-$q$. The background concentration
field $c^{\infty}(\boldsymbol{b})$ at the surface of the particle
is expanded in an analogous manner to $c(\boldsymbol{b})$, with coefficients
denoted by $\boldsymbol{C}^{\infty(q)}$. Linearity of the Laplace equation
implies that the general solution in a basis of TSH can be written
as

\begin{equation}
\boldsymbol{C}^{(q)}=\boldsymbol{\zeta}^{(q,q')}\odot\boldsymbol{C}^{\infty(q')}+\boldsymbol{\mathcal{E}}^{(q,q')}\odot\boldsymbol{J}^{(q')},\label{eq:conc-general}
\end{equation}
corresponding to Eq. (IIe), where the task now is to find the connecting
tensors $\boldsymbol{\zeta}^{(q,q')}$ and $\boldsymbol{\mathcal{E}}^{(q,q')}$.
In Appendix \ref{sec:Chemical-problem}, starting from the BIE for
the surface concentration and using a Galerkin-Jacobi iterative method,
we outline how to find approximate solutions, in leading powers of
distance between the particle and surrounding boundaries, for these
tensors in terms of a given Green's function $H$ of Laplace equation
\citep{singhCompetingChemicalHydrodynamic2019}. 

Any Green's function $H$ of Laplace equation can be written as the
sum
\begin{equation}
H(\boldsymbol{R},\tilde{\boldsymbol{R}})=H^{o}(\boldsymbol{r})+H^{*}(\boldsymbol{R},\tilde{\boldsymbol{R}}),\label{eq:Green-split}
\end{equation}
with $\boldsymbol{r}=\boldsymbol{R}-\tilde{\boldsymbol{R}}$, where
$\boldsymbol{R}$ and $\tilde{\boldsymbol{R}}$ are the field and
the source point, respectively. Here, $H^{o}(\boldsymbol{r})=1/4\pi Dr$
is the fundamental solution of Laplace equation in an unbounded domain.
On the other hand, $H^{*}$ is an extra contribution needed to satisfy
additional boundary conditions in the system. For the unbounded case,
where $H=H^{o}(\boldsymbol{r})$, the single-layer and double-layer
operators in Eqs. (IIb) and (IIc) have singular integral kernels.
However, due to translational invariance they can be evaluated using
Fourier techniques, see Appendix \ref{subsec:Exact-solution-for}.
We find that both integral operators diagonalise simultaneously in
a basis of TSH, yielding

\begin{equation}
\boldsymbol{\zeta}^{(q,q')}=\zeta_{q}\,\boldsymbol{I}^{(q,q')},\qquad\boldsymbol{\mathcal{E}}^{(q,q')}=\mathcal{E}_{q}\,\boldsymbol{I}^{(q,q')},\label{eq:unbounded-chem-exact}
\end{equation}
where
\begin{equation}
\zeta_{q}=\frac{2q+1}{q+1},\qquad\mathcal{E}_{q}=\frac{b}{D}\frac{\tilde{w}_{q}}{w_{q}}\frac{1}{q+1},\label{eq:exact-chem-coeff}
\end{equation}
and $\boldsymbol{I}^{(q,q')}$ is a tensor with elements $\delta_{qq'}$,
where the latter denotes a Kronecker delta. The expression for the
elastance $\mathcal{E}_{q}$ is confirmed by previous results obtained
by first solving the Laplace equation in the fluid volume and subsequently
matching the boundary condition (Ic) for the surface flux \citep{jackson1962classical,golestanian2007designing}.
If the system contains additional boundaries, we find corrections
to these diagonal expressions in terms of derivatives of $H^{*}$.
To leading order, this yields 

\begin{equation}
\boldsymbol{\zeta}^{(q,q')}=\zeta_{q}\left(\boldsymbol{I}^{(q,q')}+4\pi bDw_{q'}\tfrac{q'}{q'+1}b^{q+q'}\boldsymbol{\nabla}^{(q)}\tilde{\boldsymbol{\nabla}}^{(q')}H^{*}(\boldsymbol{R},\tilde{\boldsymbol{R}})\right),\label{eq:zetaqq'}
\end{equation}
where $\nabla_{\alpha_{1}\dots\alpha_{q}}^{(q)}=\nabla_{\alpha_{1}}\dots\nabla_{\alpha_{q}}$,
and where we have introduced the short-hand notation $\boldsymbol{\nabla}_{\boldsymbol{R}}=\boldsymbol{\nabla}$
for derivatives with respect to the field point and $\boldsymbol{\nabla}_{\tilde{\boldsymbol{R}}}=\tilde{\boldsymbol{\nabla}}$
for the source point. Similarly, we find for the elastance
\begin{equation}
\boldsymbol{\mathcal{E}}^{(q,q')}=\mathcal{E}_{q}\left(\boldsymbol{I}^{(q,q')}+4\pi bDw_{q}\tfrac{2q'+1}{q'+1}b^{q+q'}\boldsymbol{\nabla}^{(q)}\tilde{\boldsymbol{\nabla}}^{(q')}H^{*}(\boldsymbol{R},\tilde{\boldsymbol{R}})\right).\label{eq:epsqq'}
\end{equation}
In these expressions, the point of evaluation, $\boldsymbol{R}=\tilde{\boldsymbol{R}}$,
for the one-body problem, is left implicit for brevity.

\subsection{Hydrodynamic problem and Brownian motion\label{subsec:Slip-and-resulting}}

Using the linearity of Stokes flow we solve for the hydrodynamic traction
$\boldsymbol{f}^{H}$ in a basis of TSH. Upon eliminating the hydrodynamic
problem, Newton's equations (\ref{eq:Newton}) will reveal the Brownian
motion of an active particle. First, to find the linear system corresponding
to Eq. (IIk), we expand the slip and the hydrodynamic traction in
a basis of TSH

\begin{gather}
\boldsymbol{v}^{\mathcal{A}}(\boldsymbol{b})=\sum_{l=1}^{\infty}w_{l-1}\boldsymbol{V}^{(l)}\odot\boldsymbol{Y}^{(l-1)}(\hat{\boldsymbol{b}}),\qquad\boldsymbol{f}^{H}(\boldsymbol{b})=\sum_{l=1}^{\infty}\tilde{w}_{l-1}\boldsymbol{F}^{(l)}\odot\boldsymbol{Y}^{(l-1)}(\hat{\boldsymbol{b}}).\label{eq:exp-vs}
\end{gather}
The coefficients $\boldsymbol{V}^{(l)}$ and $\boldsymbol{F}^{(l)}$
are rank-$l$ tensors, symmetric and traceless in their last $l-1$
indices. They can be decomposed into irreducible representations,
denoted by $\boldsymbol{V}^{(l\sigma)}$ (or $\boldsymbol{F}^{(l\sigma)}$
for the traction moments), where $\boldsymbol{V}^{(ls)}$ (symmetric
and traceless), $\boldsymbol{V}^{(la)}$ (anti-symmetric) and $\boldsymbol{V}^{(lt)}$
(trace) are irreducible tensors of rank $l$, $l-1$ and $l-2$, respectively
\citep{singhManybodyMicrohydrodynamicsColloidal2015}. For slip restricted
by mass conservation only, obeying $\int\boldsymbol{v}^{\mathcal{A}}\cdot\hat{\boldsymbol{b}}{\rm d}S=0$,
these irreducible components of $\boldsymbol{V}^{(l)}$ (and $\boldsymbol{F}^{(l)}$)
are independent of each other. In terms of the common definitions
for the velocity and angular velocity of an active particle in an
unbounded domain \citep{anderson1991,stonePropulsionMicroorganismsSurface1996,ghoseIrreducibleRepresentationsOscillatory2014},

\begin{equation}
\boldsymbol{V}^{\mathcal{A}}=-\frac{1}{4\pi b^{2}}\int\boldsymbol{v}^{\mathcal{A}}(\boldsymbol{b}){\rm d}S,\qquad\boldsymbol{\Omega}^{\mathcal{A}}=-\frac{3}{8\pi b^{3}}\int\hat{\boldsymbol{b}}\times\boldsymbol{v}^{\mathcal{A}}(\boldsymbol{b}){\rm d}S,\label{eq:active-vel}
\end{equation}
we have $\boldsymbol{V}^{(1s)}=-\boldsymbol{V}^{\mathcal{A}}$ and
$\boldsymbol{V}^{(2a)}/2b=-\boldsymbol{\Omega}^{\mathcal{A}}.$ Similarly,
we have for the hydrodynamic force and torque defined in Eq. (\ref{eq:forcetorque}),
$\boldsymbol{F}^{(1s)}=\boldsymbol{F}^{H}$ and $\boldsymbol{F}^{(2a)}=\frac{1}{b}\boldsymbol{T}^{H}$.

Linearity of the Stokes equation then allows us to write down the deterministic
part of Eq. (IIk) in a basis of TSH

\begin{align}
\boldsymbol{F}^{(l\sigma)} & =-\boldsymbol{\gamma}^{(l\sigma,1s)}\cdot\boldsymbol{V}-\boldsymbol{\gamma}^{(l\sigma,2a)}\cdot\boldsymbol{\Omega}-\hat{\boldsymbol{\gamma}}^{(l\sigma,l'\sigma')}\odot\boldsymbol{V}^{(l'\sigma')},\label{eq:Stokes}
\end{align}
where $\boldsymbol{\gamma}^{(l\sigma,l'\sigma')}$ and $\hat{\boldsymbol{\gamma}}^{(l\sigma,l'\sigma')}$
are generalised friction tensors for rigid-body motion and slip, respectively.
For the modes corresponding to rigid-body motion it is known that
$\boldsymbol{\gamma}^{(l\sigma,1s)}=\hat{\boldsymbol{\gamma}}^{(l\sigma,1s)}$
and $\boldsymbol{\gamma}^{(l\sigma,2a)}=\hat{\boldsymbol{\gamma}}^{(l\sigma,2a)}$
\citep{singhGeneralizedStokesLaws2018,turkStokesTractionActive2022}.
Therefore, we can write for the hydrodynamic force and torque 

\begin{equation}
\begin{pmatrix}\boldsymbol{F}^{H}\\
\boldsymbol{T}^{H}
\end{pmatrix}=-\boldsymbol{\Gamma}\cdot\begin{pmatrix}\boldsymbol{V}-\boldsymbol{V}^{\mathcal{A}}\\
\boldsymbol{\Omega}-\boldsymbol{\Omega}^{\mathcal{A}}
\end{pmatrix}-\sum_{l\sigma=2s}\hat{\boldsymbol{\Gamma}}^{(l\sigma)}\odot\boldsymbol{V}^{(l\sigma)},\quad\text{with}\quad\boldsymbol{\Gamma}=\begin{pmatrix}\boldsymbol{\gamma}^{TT} & \boldsymbol{\gamma}^{TR}\\
\boldsymbol{\gamma}^{RT} & \boldsymbol{\gamma}^{RR}
\end{pmatrix},\quad\hat{\boldsymbol{\Gamma}}^{(l\sigma)}=\begin{pmatrix}\hat{\boldsymbol{\gamma}}^{(T,l\sigma)}\\
\hat{\boldsymbol{\gamma}}^{(R,l\sigma)}
\end{pmatrix},\label{eq:friction-pic}
\end{equation}
where the superscripts $T$ and $R$ imply $l\sigma=1s,2a$, respectively,
to confirm with existing literature \citep{laddHydrodynamicInteractionsSuspension1988}.
The matrix $\boldsymbol{\Gamma}$ contains the friction on the particle
due to rigid-body motion, and $\hat{\boldsymbol{\Gamma}}^{(l\sigma)}$
contains the friction due to higher modes of slip. This concludes
the solution of the hydrodynamic problem without fluctuations. 

In a thermally fluctuating fluid, the Brownian forces and torques
obey the fluctuation-dissipation relations \citep{einstein1905theory,zwanzig1964hydrodynamic,chowSimultaneousTranslationalRotational1973,singhFluctuatingHydrodynamicsBrownian2017}
\begin{equation}
\left\langle \begin{pmatrix}\hat{\boldsymbol{F}}(t)\\
\hat{\boldsymbol{T}}(t)
\end{pmatrix}\right\rangle =\boldsymbol{0},\qquad\left\langle \begin{pmatrix}\hat{\boldsymbol{F}}(t)\\
\hat{\boldsymbol{T}}(t)
\end{pmatrix}\begin{pmatrix}\hat{\boldsymbol{F}}(t')\\
\hat{\boldsymbol{T}}(t')
\end{pmatrix}^{\text{tr}}\right\rangle =2k_{B}T\,\boldsymbol{\Gamma}\,\delta(t-t'),\label{eq:fdt}
\end{equation}
where angled brackets denote ensemble averages, $k_{B}$ is the Boltzmann
constant and $T$ is the temperature, while the transpose is defined
as $(A_{\alpha\beta})^{\text{tr}}=A_{\beta\alpha}$. Inserting the
above equations for the deterministic and stochastic forces and torques
into Newton's equations (\ref{eq:Newton}) yields the Langevin equation
\begin{equation}
\begin{pmatrix}m\dot{\boldsymbol{V}}\\
I\dot{\boldsymbol{\Omega}}
\end{pmatrix}=\begin{pmatrix}\boldsymbol{F}^{P}\\
\boldsymbol{T}^{P}
\end{pmatrix}-\boldsymbol{\Gamma}\cdot\begin{pmatrix}\boldsymbol{V}-\boldsymbol{V}^{\mathcal{A}}\\
\boldsymbol{\Omega}-\boldsymbol{\Omega}^{\mathcal{A}}
\end{pmatrix}-\sum_{l\sigma=2s}\hat{\boldsymbol{\Gamma}}^{(l\sigma)}\odot\boldsymbol{V}^{(l\sigma)}+\sqrt{2k_{B}T\,\boldsymbol{\Gamma}}\cdot\begin{pmatrix}\boldsymbol{\xi}^{T}\\
\boldsymbol{\xi}^{R}
\end{pmatrix}.\label{eq:Newton-FDT}
\end{equation}
The parameters $\boldsymbol{\xi}^{\alpha}$ are random variables with zero mean
and unit variance. In the inertial equation (\ref{eq:Newton-FDT})
the noise is not multiplicative since $\boldsymbol{\Gamma}$ is configuration
dependent, but not velocity dependent. With the particle centre of
mass $\boldsymbol{R}$ and its unit orientation vector $\boldsymbol{e}$
(its orientation is governed by the rotational dynamics $\dot{\boldsymbol{\Theta}}=\boldsymbol{\Omega}$,
where $\boldsymbol{\Theta}$ is an arbitrary set of angles), we can
find its Brownian trajectory by integrating
\begin{equation}
\dot{\boldsymbol{R}}=\boldsymbol{V},\qquad\dot{\boldsymbol{e}}=\boldsymbol{\Omega}\times\boldsymbol{e},\label{eq:dynamics}
\end{equation}
over time. In colloidal systems the inertia of both the particles
and the fluid are typically negligible. This corresponds to the Smoluchowski
limit of Eq. (\ref{eq:Newton-FDT}). Adiabatic elimination of the
momentum variables in phase space then directly leads to the following
update equations in It\^{o} form \citep{ermakBrownianDynamicsHydrodynamic1978,gardiner1984adiabatic,wajnrybBrownianDynamicsDivergence2004,volpeEffectiveDriftsDynamical2016}:
\begin{subequations}\label{eq:EoM}
\begin{align}
\boldsymbol{R}(t+\Delta t) & =\boldsymbol{R}(t)+\Big\{\boldsymbol{V}^{\mathcal{A}}+\boldsymbol{\mu}^{TT}\cdot\boldsymbol{F}^{P}+\boldsymbol{\mu}^{TR}\cdot\boldsymbol{T}^{P}+\sum_{l\sigma=2s}\boldsymbol{\pi}^{(T,l\sigma)}\odot\boldsymbol{V}^{(l\sigma)}+k_{B}T\,\tilde{\boldsymbol{\nabla}}\cdot\boldsymbol{\mu}^{TT}\Big\}\Delta t+\Delta\hat{\boldsymbol{R}},\label{eq:update-position}\\
\nonumber \\
\boldsymbol{e}(t+\Delta t) & =\boldsymbol{e}(t)+\Big\{\boldsymbol{\Omega}^{\mathcal{A}}+\boldsymbol{\mu}^{RT}\cdot\boldsymbol{F}^{P}+\boldsymbol{\mu}^{RR}\cdot\boldsymbol{T}^{P}+\sum_{l\sigma=2s}\boldsymbol{\pi}^{(R,l\sigma)}\odot\boldsymbol{V}^{(l\sigma)}+k_{B}T\,\tilde{\boldsymbol{\nabla}}\cdot\boldsymbol{\mu}^{RT}\Big\}\Delta t\times\boldsymbol{e}(t)+\Delta\hat{\boldsymbol{e}},\label{eq:update-orientation}
\end{align}
\end{subequations}with $\Delta\hat{\boldsymbol{e}}=\Delta\hat{\boldsymbol{\Theta}}(t)\times\boldsymbol{e}(t)+\tfrac{1}{2}\Delta\hat{\boldsymbol{\Theta}}(t)\cdot\left[\boldsymbol{e}(t)\Delta\hat{\boldsymbol{\Theta}}(t)-\Delta\hat{\boldsymbol{\Theta}}(t)\boldsymbol{e}(t)\right]$,
while
\begin{equation}
\left\langle \begin{pmatrix}\Delta\hat{\boldsymbol{R}}\\
\Delta\hat{\boldsymbol{\Theta}}
\end{pmatrix}\right\rangle =\boldsymbol{0},\qquad\left\langle \begin{pmatrix}\Delta\hat{\boldsymbol{R}}\\
\Delta\hat{\boldsymbol{\Theta}}
\end{pmatrix}\begin{pmatrix}\Delta\hat{\boldsymbol{R}}\\
\Delta\hat{\boldsymbol{\Theta}}
\end{pmatrix}^{\text{tr}}\right\rangle =2k_{B}T\,\mathbb{M}\,\Delta t.\label{eq:update-error}
\end{equation}
It is clear that the grand mobility matrix $\mathbb{M}$ and the
grand propulsion tensor $\boldsymbol{\Pi}^{(l\sigma)}$ satisfy
\begin{equation}
\mathbb{M}=\begin{pmatrix}\boldsymbol{\mu}^{TT} & \boldsymbol{\mu}^{TR}\\
\boldsymbol{\mu}^{RT} & \boldsymbol{\mu}^{RR}
\end{pmatrix}=\boldsymbol{\Gamma}^{-1},\qquad\boldsymbol{\Pi}^{(l\sigma)}=\begin{pmatrix}\boldsymbol{\pi}^{(T,l\sigma)}\\
\boldsymbol{\pi}^{(R,l\sigma)}
\end{pmatrix}=-\mathbb{M}\cdot\hat{\boldsymbol{\Gamma}}^{(l\sigma)}.
\end{equation}
Onsager-Casimir symmetry implies symmetry of the mobility matrix,
and we can identify the so-called propulsion tensors as $\boldsymbol{\pi}^{(\alpha,l\sigma)}=-\boldsymbol{\mu}^{\alpha T}\cdot\hat{\boldsymbol{\gamma}}^{(T,l\sigma)}-\boldsymbol{\mu}^{\alpha R}\cdot\hat{\boldsymbol{\gamma}}^{(R,l\sigma)},$
with $\alpha\in\{T,R\}$ \citep{singhGeneralizedStokesLaws2018}.
The convective terms in the update equations constitute the thermal
drift, which arises from a simple forward Euler integration scheme
of the Langevin equations. The occurring derivative $\tilde{\boldsymbol{\nabla}}$
is the standard spatial gradient (with respect to the source point).
If the mobilities depend on the particle orientation, additional orientational
convective terms must be included. For the spherical particles considered
here, however, these terms do not contribute. The quadratic term in
$\Delta\boldsymbol{\Theta}$ in $\Delta\hat{\boldsymbol{e}}$ is needed
to preserve the condition $|\boldsymbol{e}|=1$ as discussed in \citep{makinoBrownianMotionParticle2004,decoratoHydrodynamicsBrownianMotions2015}. 

As the Stokes equation defines a dissipative system, any acceptable approximation
of $\mathbb{M}$ must remain positive-definite for all physical configurations,
e.g. when a simulated particle does not overlap with nearby boundaries
\citep{cichockiFrictionMobilityColloidal2000}. In Appendix \ref{sec:Hydrodynamic-problem--},
starting from the BIE of Stokes flow and using a Galerkin-Jacobi iterative
method, we outline how to find such solutions, in principle to arbitrary
accuracy in the distance between the particle and surrounding boundaries,
for the mobility and propulsion tensors in terms of the Green's function
$\boldsymbol{G}$ of Stokes flow. For this, we write the Green's function
as the sum \citep{mariansmoluchowskiUeberWechselwirkungKugeln1911},
\begin{equation}
\boldsymbol{G}(\boldsymbol{R},\tilde{\boldsymbol{R}})=\boldsymbol{G}^{o}(\boldsymbol{r})+\boldsymbol{G}^{*}(\boldsymbol{R},\tilde{\boldsymbol{R}}),\label{eq:oseenplus}
\end{equation}
where $\boldsymbol{r}=\boldsymbol{R}-\tilde{\boldsymbol{R}}$, and
$\boldsymbol{G}^{o}(\boldsymbol{r})=\left(\boldsymbol{I}+\hat{\boldsymbol{r}}\hat{\boldsymbol{r}}\right)/8\pi\eta r$
is the Oseen tensor for unbounded Stokes flow \citep{oseenHydrodynamik1927,pozrikidisBoundaryIntegralSingularity1992}.
The term $\boldsymbol{G}^{*}$ is the correction necessary to satisfy
additional boundary conditions in the system. In the unbounded domain,
where $\boldsymbol{G}=\boldsymbol{G}^{o}(\boldsymbol{r})$, the mobility
matrix $\mathbb{M}$ diagonalises and the propulsion tensors vanish
identically,
\begin{equation}
\boldsymbol{\mu}^{TT}=\mu_{T}\,\boldsymbol{I},\quad\boldsymbol{\mu}^{R}=\mu_{R}\,\boldsymbol{I},\quad\boldsymbol{\mu}^{TR}=\boldsymbol{\mu}^{RT}=\boldsymbol{0},\qquad\boldsymbol{\pi}^{(\alpha,l\sigma)}=\boldsymbol{0}.
\end{equation}
Here, $\mu_{T}=\left(6\pi\eta b\right)^{-1}$ and $\mu_{R}=\left(8\pi\eta b^{3}\right)^{-1}$
are the well-known mobility coefficients for translation and rotation
of a sphere of radius $b$ in an unbounded fluid of viscosity $\eta$
\citep{stokesEffectInternalFriction1850a}. For a system containing
additional boundaries, we obtain corrections to the above expressions
in terms of derivatives of $\boldsymbol{G}^{*}$. As shown in the
Appendix, to leading order in the Jacobi iteration the mobilities
are
\begin{equation}\label{eq:mobilityleading}
\boldsymbol{\mu}^{TT}=\mu_{T}\left(\boldsymbol{I}+6\pi\eta b\,\mathcal{F}^{0}\tilde{\mathcal{F}}^{0}\boldsymbol{G}^{*}\right),\qquad\boldsymbol{\mu}^{TR}=\tfrac{1}{2}\mathcal{F}^{0}\,\tilde{\boldsymbol{\nabla}}\times\boldsymbol{G}^{*},\qquad\boldsymbol{\mu}^{RR}=\mu_{R}\,\boldsymbol{I}+\tfrac{1}{2}\boldsymbol{\nabla}\times\boldsymbol{\mu}^{TR},
\end{equation}
where we have defined the differential
operators $\mathcal{F}^{l}=\big(1+\frac{b^{2}}{4l+6}\nabla^{2}\big)$
and $\tilde{\mathcal{F}}^{l}=\big(1+\frac{b^{2}}{4l+6}\tilde{\nabla}^{2}\big)$.

Governed by the particle's activity, we choose to retain the leading
symmetric and polar modes of the slip. As demonstrated in the next
Section, this requires the following propulsion tensors:
\begin{equation}\label{eq:propulsionleading}
\boldsymbol{\pi}^{(T,2s)}=\tfrac{10\pi\eta b^{2}}{3}\mathcal{F}^{0}\tilde{\mathcal{F}}^{1}\left[\tilde{\boldsymbol{\nabla}}\boldsymbol{G}^{*}+(\tilde{\boldsymbol{\nabla}}\boldsymbol{G}^{*})^{\text{tr}}\right],\qquad\boldsymbol{\pi}^{(T,3t)}=-\tfrac{2\pi\eta b^{3}}{5}\mathcal{F}^{0}\tilde{\nabla}^{2}\boldsymbol{G}^{*},\qquad\boldsymbol{\pi}^{(T,4t)}=-\tfrac{2\pi\eta b^{4}}{63}\mathcal{F}^{0}\tilde{\boldsymbol{\nabla}}\tilde{\nabla}^{2}\boldsymbol{G}^{*},
\end{equation}
given to leading order in the Jacobi
iteration. The structure of the problem implies that $\boldsymbol{\pi}^{(R,l\sigma)}=\tfrac{1}{2}\left(\boldsymbol{\nabla}\times\boldsymbol{\pi}^{(T,l\sigma)}\right).$
To the given order these have been first obtained in Ref. \citep{singhGeneralizedStokesLaws2018}. 

\subsection{Chemo-hydrodynamic coupling and resulting propulsion\label{subsec:Chemohydrodynamic-coupling}}

We now consider the boundary condition (IIl), coupling the hydrodynamic
to the chemical problem. We observe that the differential operator
$\boldsymbol{\chi}$ defined in Eq. (Ig) implies tangential slip such
that $\hat{\boldsymbol{b}}\cdot\boldsymbol{v}^{\mathcal{A}}=0$, i.e.,
chemical gradients at the surface of the particle can only drive tangential
slip flows. Satisfying this condition, we write the tangential modes
in the expansion of the slip in Eq. (\ref{eq:exp-vs}) with a subscript
$s$ as $\boldsymbol{V}_{s}^{(l\sigma)}$. In order to obey the tangential
slip condition, the symmetric and trace modes of the slip expansion
coefficients have to satisfy 
\begin{equation}
\boldsymbol{V}_{s}^{([l+2]t)}=-l(2l+3)\boldsymbol{V}_{s}^{(ls)}.\label{eq:mode-locking}
\end{equation}
This means that, whenever a $\boldsymbol{V}_{s}^{(ls)}$ mode is generated,
a $\boldsymbol{V}_{s}^{([l+2]t)}$ mode of strength given by Eq. (\ref{eq:mode-locking})
will be generated too. For the anti-symmetric modes $\boldsymbol{V}_{s}^{(la)}$
there is no such condition as they produce tangential slip flow by
definition \citep{singhManybodyMicrohydrodynamicsColloidal2015}.

Finally, to express the boundary condition (Ig) in a basis of TSH,
we expand the phoretic mobility as

\begin{equation}
\mu_{c}(\boldsymbol{b})=\sum_{q=0}^{\infty}\tilde{w}_{q}\boldsymbol{M}^{(q)}\odot\boldsymbol{Y}^{(q)}(\hat{\boldsymbol{b}}).\label{eq:exp-m}
\end{equation}
The coefficients $\boldsymbol{M}^{(q)}$ are symmetric and traceless
tensors of rank-$q$. This yields the linear system corresponding
to Eq. (IIl),
\begin{align}
\boldsymbol{V}_{s}^{(l\sigma)} & =\boldsymbol{\chi}^{(l\sigma,q)}\odot\boldsymbol{C}^{(q)}.\label{eq:slip-modes}
\end{align}
The coupling tensor $\boldsymbol{\chi}^{(l\sigma,q)}$ is given in
Appendix \ref{sec:Chemohydrodynamic-coupling}, and satisfies $\boldsymbol{\chi}^{(l\sigma,0)}=\boldsymbol{0}$,
i.e., a uniform surface concentration does not induce slip. 

In principle, any form of tangential slip can be generated by the
chemo-hydrodynamic coupling in Eq. (\ref{eq:slip-modes}). Here, we
only consider the leading polar ($\boldsymbol{V}_{s}^{(3t)}$), chiral
($\boldsymbol{V}_{s}^{(2a)}$) and symmetric ($\boldsymbol{V}_{s}^{(2s)}$)
modes. Using Eq. (\ref{eq:mode-locking}), we can identify

\begin{equation}
\boldsymbol{V}^{\mathcal{A}}=-\boldsymbol{V}_{s}^{(1s)}=\tfrac{1}{5}\boldsymbol{V}_{s}^{(3t)},\qquad2b\boldsymbol{\Omega}^{\mathcal{A}}=-\boldsymbol{V}_{s}^{(2a)},\qquad\boldsymbol{V}_{s}^{(2s)}=-\tfrac{1}{14}\boldsymbol{V}_{s}^{(4t)}.\label{eq:modes}
\end{equation}
In the following, we therefore parametrise polar, chiral and symmetric
slip by $\boldsymbol{V}^{\mathcal{A}}$, $\boldsymbol{\Omega}^{\mathcal{A}}$
and $\boldsymbol{V}_{s}^{(2s)}$, respectively. With this, the propulsion
terms in the update equations (\ref{eq:EoM}) are
\begin{equation}
\sum_{l\sigma=2s}\boldsymbol{\Pi}^{(l\sigma)}\odot\boldsymbol{V}^{(l\sigma)}=5\begin{pmatrix}\boldsymbol{\pi}^{(T,3t)}\\
\boldsymbol{\pi}^{(R,3t)}
\end{pmatrix}\cdot\boldsymbol{V}^{\mathcal{A}}+\begin{pmatrix}\boldsymbol{\pi}^{(T,2s)}-14\boldsymbol{\pi}^{(T,4t)}\\
\boldsymbol{\pi}^{(R,2s)}-14\boldsymbol{\pi}^{(R,4t)}
\end{pmatrix}\colon\boldsymbol{V}_{s}^{(2s)}\label{eq:prop-terms}
\end{equation}
for an autophoretic particle, and Eq. (\ref{eq:slip-modes}) yields\begin{subequations}\label{eq:rel-slip-modes}\begingroup\setlength{\jot}{1ex}
\begin{align}
\boldsymbol{V}^{\mathcal{A}} & =-\tfrac{1}{4\pi b^{3}}\sum_{q=1}^{\infty}\left[\tfrac{q+1}{2q+1}\boldsymbol{M}^{(q-1)}-q(q+1)\boldsymbol{M}^{(q+1)}\right]\odot\boldsymbol{C}^{(q)},\\
\boldsymbol{\Omega}^{\mathcal{A}} & =-\tfrac{3}{8\pi b^{4}}\sum_{q=1}^{\infty}q\boldsymbol{M}^{(q)}\times'\boldsymbol{C}^{(q)},\\
\boldsymbol{V}_{s}^{(2s)} & =\tfrac{3}{4\pi b^{3}}\sum_{q=1}^{\infty}\left[\tfrac{q+1}{4q^{2}-1}\boldsymbol{M}^{(q-2)}+\tfrac{3q}{2q+3}\boldsymbol{M}_{{\rm sym}}^{(q)}-q(q+1)(q+2)\boldsymbol{M}^{(q+2)}\right]\odot\boldsymbol{C}^{(q)}.
\end{align}
\endgroup\end{subequations}For brevity, we have left the solution
for $\boldsymbol{C}^{(q)}$ of Eq. (\ref{eq:conc-general}) implicit.
Here, we have defined a cross-product for irreducible tensors as $(\boldsymbol{M}^{(q)}\times'\boldsymbol{C}^{(q)}){}_{\alpha}=\epsilon_{\alpha\beta\gamma}M_{\beta({\scriptscriptstyle Q-1})}^{(q)}C_{\gamma({\scriptscriptstyle Q-1})}^{(q)}$
and a symmetric and traceless product contracting $(q-1)$ indices,
$(\boldsymbol{M}_{{\rm sym}}^{(q)}\odot\boldsymbol{C}^{(q)})_{\alpha\beta}=\Delta_{\alpha\beta,\alpha'\beta'}^{(2)}M_{\alpha'({\scriptscriptstyle Q-1})}^{(q)}C_{\beta'({\scriptscriptstyle Q-1})}^{(q)}$,
where we have used the short-hand notation $Q=\gamma_{1}\gamma_{2}\dots\gamma_{q}$
for Cartesian indices \citep{damourMultipoleAnalysisElectromagnetism1991}.

\section{Applications\label{sec:Applications}}

In this section, we demonstrate the methodology introduced thus far
with the help of three examples. First, we discuss model design to
achieve certain types of motion and the effect of thermal fluctuations
in the bulk fluid. With the help of the appropriate Green's functions,
we then provide the chemical and hydrodynamic connectors necessary
to describe the dynamics of a phoretic particle near a plane, chemically
permeable surface of two immiscible liquids. In a representative example,
we investigate some of the chemo-hydrodynamic effects this interface
has on the motion of a self-rotating autophoretic particle. Finally,
we discuss the hovering state of an isotropic chemical source particle
above an interface as a function of particle activity, and the chemo-hydrodynamic
properties of the interface.

We use the following notation for the uniaxial parameterisation of
the $q$-th modes of the phoretic mobility and surface flux:
\begin{equation}
\bm{M}^{(q)}=M_{q}\bm{Y}^{(q)}(\bm{p}_{q}),\qquad\bm{J}^{(q)}=J_{q}\bm{Y}^{(q)}(\bm{e}_{q}).\label{eq:trunc-chem}
\end{equation}
Here, $M_{q}$ and $J_{q}$ are constants representing the strength
of the $q$-th mode, while $\bm{p}_{q}$ and $\bm{e}_{q}$ are unit
vectors. 

\subsection{Programmed Brownian motion in the bulk\label{subsec:bulk}}

In the bulk, far away from boundaries, we can simplify Eqs. (\ref{eq:EoM}).
First, we non-dimensionalise the equations by rescaling velocities
by the speed of a particle with constant phoretic mobility, namely
$4\pi b^{2}\,V=\mu_{c}J_{1}/D$. Angular velocities are rescaled by
$V/b$. Renaming rescaled variables such that Eq. (\ref{eq:dynamics})
reads the same, we obtain
\begin{equation}
\partial_{t}\bm{R}=\bm{V}^{\mathcal{A}}+\sqrt{2D_{T}}\,\bm{\xi}_{T},\qquad\partial_{t}\bm{e}=\left(\bm{\Omega}^{\mathcal{A}}+\sqrt{2D_{R}}\,\bm{\xi}_{R}\right)\times\bm{e},\label{eq:EoM-stoch}
\end{equation}
where for a spherical body in an unbounded fluid the translational
diffusivity $D_{T}=\mathcal{B}/6$ and the rotational diffusivity
$D_{R}=\mathcal{B}/8$ are isotropic and defined in terms of the Brown
number
\begin{equation}
\mathcal{B}=\frac{k_{B}T}{\pi\eta b^{2}V},
\end{equation}
the ratio of Brownian to hydrodynamic forces. Analogously, a particle
P\'{e}clet number can be defined by ${\rm Pe}=1/\mathcal{B}$ \citep{mozaffariSelfpropelledColloidalParticle2018}.
For a model including modes up to second order in both the phoretic
mobility and the flux expansion, the velocity and angular velocity
read\begin{subequations}
\begin{align}
\bm{V}^{\mathcal{A}} & =-\bm{e}_{1}+3m_{2}\left[3\bm{p}_{2}\left(\bm{p}_{2}\cdot\bm{e}_{1}\right)-\bm{e}_{1}\right]-6m_{1}j_{2}\left[3\bm{e}_{2}\left(\bm{e}_{2}\cdot\bm{p}_{1}\right)-\bm{p}_{1}\right],\\
\bm{\Omega}^{\mathcal{A}} & =-\tfrac{9}{4}m_{1}\left(\bm{p}_{1}\times\bm{e}_{1}\right)-270m_{2}j_{2}\left(\bm{p}_{2}\cdot\bm{e}_{2}\right)\left(\bm{p}_{2}\times\bm{e}_{2}\right).
\end{align}
\end{subequations}where $m_{i}=M_{i}/M_{0}$ and $j_{2}=J_{2}/J_{1}$.
As will be convenient later, we define the angles $\bm{e}_{1}\cdot\bm{p}_{i}=\cos\alpha_{i}$,
$\bm{e}_{1}\cdot\bm{e}_{2}=\cos\beta$ and $\bm{e}_{2}\cdot\bm{p}_{2}=\cos\gamma$.
Without loss of generality, we choose $\bm{e}_{1}$ as the orientation
of the particle. This constitutes a minimal model capable of modelling
the five distinct types of motion \citep{lisickiAutophoreticMotionThree2018}:
1) pure translation, 2) pure rotation (spinning), 3) parallel rotation
and translation, 4) perpendicular rotation and translation (circular
swimming), and 5) helical motion. In the following we briefly discuss
particle designs for each type of motion and analyse how thermal noise
affects the dynamics by computing the mean-squared displacement of
selected examples. 

\begin{figure}
\centering
\includegraphics[width=\columnwidth]{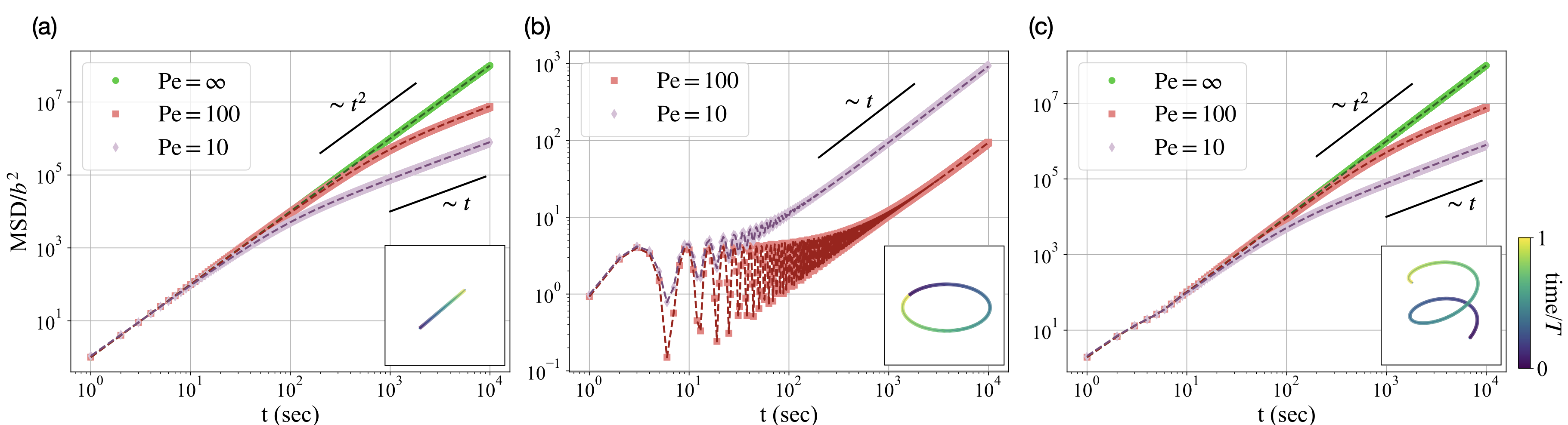}
\vspace{-4mm}\caption{\label{fig:msd}\textbf{Mean-squared displacement of the programmed
Brownian motion of an autophoretic particle.} In panels (a), (b),
and (c) we compare the non-dimensionalised MSDs of translational,
circular and helical swimming computed for Brownian simulations with
their theoretical predictions (see the main text for the latter) at
various temperatures characterised by a particle P\'{e}clet number
${\rm Pe}$. We define ${\rm Pe}=\infty$ (no noise), ${\rm Pe}=100$
(moderate noise) and ${\rm Pe}=10$ (strong noise) following Mozaffari
et al. \citep{mozaffariSelfpropelledColloidalParticle2018}. For all
three types of motion a diffusive regime $\sim t$ can be identified
above a certain persistence time broadly determined by the amount
of rotational diffusivity. The insets show the respective trajectories
over an arbitrary time $T$ in the limit of zero temperature.}
\end{figure}
Pure translation is the simplest kind of motion and is achieved by
choosing $m_{1}=m_{2}=j_{2}=0$. The update equations are those of
an active Brownian particle (ABP) with swimming direction $-\bm{e}_{1}$,\begin{subequations}
\begin{align}
\boldsymbol{R}(t+\Delta t) & =\boldsymbol{R}(t)-\bm{e}_{1}\Delta t+\sqrt{2D_{T}}\Delta\bm{W}_{T},\\
\boldsymbol{e}_{1}(t+\Delta t) & =\boldsymbol{e}_{1}(t)+\sqrt{2D_{R}}\Delta\bm{W}_{R}\times\boldsymbol{e}_{1}(t),
\end{align}
\end{subequations}where $\Delta\bm{W}_{T}$ and $\Delta\bm{W}_{R}$
are increments of mutually independent Wiener processes \citep{gardiner1985handbook}.
In Figure \ref{fig:msd}(a), we show the simulated (markers) and theoretical
(dashed lines) MSD for such a particle at various temperatures. The
MSD for an ABP is known exactly and is given by \citep{fodorStatisticalPhysicsActive2018}
\begin{equation}
\left\langle \Delta\bm{r}^{2}(t)\right\rangle _{{\rm tra}}=6\left(D_{T}+D_{A}\right)t+2(V\tau)^{2}\left(e^{-t/\tau}-1\right),
\end{equation}
where $D_{A}=\mu_{T}V^{2}\tau/3$ is the active diffusion coefficient
and $\tau^{-1}=2D_{R}$ is the persistence time due to rotational
noise. The persistence time indicates a transition from a ballistic
to a diffusive dynamics, clearly visible in the figure. In the limit
of zero temperature, i.e., $\mathcal{B}=0$, the MSD reduces to 
\begin{equation}
\left\langle \Delta\bm{r}^{2}(t)\right\rangle _{{\rm tra}}^{\mathcal{B}=0}=(Vt)^{2},
\end{equation}
as indicated in the figure. 

A spinning particle (pure rotation) can be modelled by choosing $\alpha_{2}=\pi/2$,
$m_{2}=-1/3$ and $j_{2}=0$ while $\sin\alpha_{1}\neq0$. This is
captured by the update equations
\begin{subequations}
\begin{align}
\boldsymbol{R}(t+\Delta t) & =\boldsymbol{R}(t)+\sqrt{2D_{T}}\Delta\bm{W}_{T},\\
\boldsymbol{e}_{1}(t+\Delta t) & =\boldsymbol{e}_{1}(t)+\tfrac{9}{4}m_{1}\left(\bm{p}_{1}(t)-\bm{e}_{1}(t)\cos\alpha_{1}\right)\Delta t+\sqrt{2D_{R}}\Delta\bm{W}_{R}\times\boldsymbol{e}_{1}(t).
\end{align}
\end{subequations}

Additional translation parallel to rotation on the other hand occurs
for the parameter values $\alpha_{1}=0$, $\alpha_{2}=\beta=\pi/2$
and $\sin(2\gamma)\neq0$, with $\bm{e}_{1}$ as the translation and
rotation axis of the update equations
\begin{subequations}
\begin{align}
\boldsymbol{R}(t+\Delta t) & =\boldsymbol{R}(t)-\bm{e}_{1}\left[1+3m_{2}-6m_{1}j_{2}\right]\Delta t+\sqrt{2D_{T}}\Delta\bm{W}_{T},\\
\boldsymbol{e}_{1}(t+\Delta t) & =\boldsymbol{e}_{1}(t)+\sqrt{2D_{R}}\Delta\bm{W}_{R}\times\boldsymbol{e}_{1}(t).
\end{align}
\end{subequations}

Circular swimming (perpendicular rotation and translation) is obtained
by choosing $m_{2}=j_{2}=0$ and $\sin\alpha_{1}\neq0$. For such
a self-rotating circle swimmer one can compute the MSD exactly if
the Brownian motion is confined to the plane perpendicular to $\bm{\Omega}^{\mathcal{A}}$
\citep{vanteeffelenDynamicsBrownianCircle2008}, 
\begin{gather}
\left\langle \Delta\bm{r}^{2}(t)\right\rangle _{{\rm circ}}=2\lambda^{2}\left\{ \Omega^{2}-D_{R}^{2}+D_{R}(D_{R}^{2}+\Omega^{2})t+e^{-D_{R}t}\left[(D_{R}^{2}-\Omega^{2})\cos\Omega t-2D_{R}\Omega\sin\Omega t\right]\right\} +4D_{T}t,
\end{gather}
where $\lambda=V/(D_{R}^{2}+\Omega^{2})$. Here, $\Omega$ represents
the circular frequency. In the limit of zero temperature, this reduces
to
\begin{equation}
\left\langle \Delta\bm{r}^{2}(t)\right\rangle _{{\rm circ}}^{\mathcal{B}=0}=2\left(\frac{V}{\Omega}\right)^{2}(1-\cos\Omega t),
\end{equation}
where $V/\Omega$ is the radius of the circular motion. Since in this
case we can restrict our attention to the $x$-$z$ plane, we can
define the planar polar angle $\vartheta$ such that $\boldsymbol{e}_{1}=\cos\vartheta\,\hat{\boldsymbol{x}}+\sin\vartheta\,\hat{\boldsymbol{z}}.$
The update equations then take the simplified form \begin{subequations}
\begin{align}
x(t+\Delta t) & =x(t)-\cos\vartheta\Delta t+\sqrt{2D_{T}}\Delta W_{x},\\
z(t+\Delta t) & =z(t)-\sin\vartheta\Delta t+\sqrt{2D_{T}}\Delta W_{z},\\
\vartheta(t+\Delta t) & =\vartheta(t)+\tfrac{9}{4}m_{1}\sin\alpha_{1}\Delta t+\sqrt{2D_{R}}\Delta W_{\vartheta},
\end{align}
\end{subequations}where $\Delta W_{x}$, $\Delta W_{z}$ and $\Delta W_{\vartheta}$
are increments of mutually independent Wiener processes. In Figure
\ref{fig:msd}(b) we compare the MSD obtained from simulations with
the theoretical expressions.

Helical motion occurs for all other parameter values, representing
a general non-axisymmetric phoretic particle. A simple example is
given by choosing $j_{2}=0$ and $\cos\beta\neq0$. The corresponding
update equations are \begin{subequations}
\begin{align}
\boldsymbol{R}(t+\Delta t) & =\boldsymbol{R}(t)+\left(9m_{2}\bm{p}_{2}\cos\beta-\bm{e}_{1}\left(1+3m_{2}\right)\right)\Delta t+\sqrt{2D_{T}}\Delta\bm{W}_{T},\\
\boldsymbol{e}_{1}(t+\Delta t) & =\boldsymbol{e}_{1}(t)+\tfrac{9}{4}m_{1}\left(\bm{p}_{1}(t)-\bm{e}_{1}(t)\cos\alpha_{1}\right)\Delta t+\sqrt{2D_{R}}\Delta\bm{W}_{R}\times\boldsymbol{e}_{1}(t).
\end{align}
\end{subequations}At zero temperature, the pitch angle $\psi$ of
the resulting helix is given by the simple expression
\begin{equation}
\frac{\bm{V}^{\mathcal{A}}}{\left|\bm{V}^{\mathcal{A}}\right|}\cdot\frac{\bm{\Omega}^{\mathcal{A}}}{\left|\bm{\Omega}^{\mathcal{A}}\right|}=-\frac{9m_{2}\sin(2\beta)}{2\sqrt{1+m_{2}(6-\cos^{2}\beta)+3m_{2}^{2}(3-26\cos^{2}\beta)}}=\cos\psi.
\end{equation}
In Figure\ref{fig:msd}(c) we approximate the corresponding MSD by
a superposition of translational and circular Brownian motion discussed
in the previous paragraphs such that
\begin{equation}
\left\langle \Delta\bm{r}^{2}(t)\right\rangle _{{\rm hel}}=\left\langle \Delta\bm{r}^{2}(t)\right\rangle _{{\rm tra}}^{V_{\perp}}+\left\langle \Delta\bm{r}^{2}(t)\right\rangle _{{\rm circ}}^{V_{\parallel}}-4D_{T}t,
\end{equation}
where the last term is introduced to avoid accounting for translational
noise twice. The superscripts of the MSD terms indicate which component
of the velocity enters the respective terms. Here, $V_{\perp}$ is
the component of the velocity perpendicular to the plane of the circular
motion, and $V_{\parallel}$ the component within that plane. Naturally,
in the limit of zero temperature this reduces to the exact deterministic
MSD for a helix
\begin{equation}
\left\langle \Delta\bm{r}^{2}(t)\right\rangle _{{\rm hel}}^{{\rm det}}=(V_{\perp}t)^{2}+2\left(\frac{V_{\parallel}}{\Omega}\right)^{2}(1-\cos\Omega t).
\end{equation}
There is good agreement between this approximation and the MSD computed
from simulated Brownian trajectories. Compared with the MSD of an ABP
in Figure\ref{fig:msd}(a) a kink indicating the period $2\pi/\Omega$
of the circular part of the motion is clearly visible. 

\subsection{Autophoresis near a permeable interface\label{subsec:interface-effect}}

\begin{table*}
\begin{tabular}{>{\centering}p{2cm}>{\raggedright}p{5.5cm}>{\raggedright}p{10cm}}
\toprule 
\addlinespace
\textbf{Region} & \textbf{Chemical Green's function} & \textbf{Hydrodynamic Green's function}\tabularnewline\addlinespace
\midrule
\midrule 
\addlinespace
$z>0$ & $\quad H(\boldsymbol{R},\tilde{\boldsymbol{R}})=H^{o}(\boldsymbol{r})+\tfrac{1-\lambda^{c}}{1+\lambda^{c}}H^{o}(\boldsymbol{r}^{*})$ & $\,\,G_{\alpha\beta}(\boldsymbol{R},\tilde{\boldsymbol{R}})=G_{\alpha\beta}^{o}(\boldsymbol{r})+\mathcal{M}_{\beta\gamma}^{f}G_{\alpha\gamma}^{o}(\boldsymbol{r}^{*})-2h\frac{\lambda^{f}}{1+\lambda^{f}}\nabla_{\gamma}^{*}G_{\alpha z}^{o}(\boldsymbol{r}^{*})\mathcal{M}_{\beta\gamma}$\tabularnewline\addlinespace
\addlinespace
 &  & \hfill{} $+h^{2}\frac{\lambda^{f}}{1+\lambda^{f}}\nabla^{*2}G_{\alpha\gamma}^{o}(\boldsymbol{r}^{*})\mathcal{M}_{\beta\gamma}$\tabularnewline\addlinespace
\midrule 
\addlinespace
$z<0$ & $\quad H(\boldsymbol{R},\tilde{\boldsymbol{R}})=\tfrac{2\lambda^{c}}{1+\lambda^{c}}H^{o}(\boldsymbol{r})$ & $\,\,G_{\alpha\beta}(\boldsymbol{R},\tilde{\boldsymbol{R}})=\frac{\lambda^{f}}{1+\lambda^{f}}\left[2\delta_{\beta\rho}G_{\alpha\rho}^{o}(\boldsymbol{r})-2h\nabla_{\beta}G_{\alpha z}^{o}(\boldsymbol{r})-h^{2}\nabla^{2}G_{\alpha\beta}^{o}(\boldsymbol{r})\right]$\tabularnewline\addlinespace
\bottomrule
\end{tabular}

\caption{\label{tab:GREEN}\textbf{Green's functions for a plane interface.}
The Green's functions for the concentration (Laplace equation, left)
and velocity (Stokes equation, right) fields near a plane, chemically
permeable fluid-fluid interface of solute diffusivity ratio $\lambda^{c}=D_{2}/D_{1}$
and viscosity ratio $\lambda^{f}=\eta_{2}/\eta_{1}$ at $z=0$ are
given. The particle is in the positive half-space $z>0$, where the
solute diffusivity is $D_{1}$ and the fluid viscosity is $\eta_{1}$.
We use $\boldsymbol{r}=\boldsymbol{R}-\tilde{\boldsymbol{R}}$ and
$\boldsymbol{r}^{*}=\boldsymbol{R}-\tilde{\boldsymbol{R}}^{*}$, with
the field point $\boldsymbol{R}=\left(x,y,z\right)^{\text{tr}}$ ,
the source point $\tilde{\boldsymbol{R}}=(\tilde{x},\tilde{y},\tilde{z})^{\text{tr}}$
and the relation $\tilde{\boldsymbol{R}}^{*}=\boldsymbol{\mathcal{M}}\cdot\tilde{\boldsymbol{R}}$
between the physical position vector $\tilde{\boldsymbol{R}}$ and
the position vector of the image singularities $\tilde{\boldsymbol{R}}^{*}$.
The height of the centre of the particle above the interface is $\tilde{z}=h$.
With this, we have $\tilde{\boldsymbol{R}}-\tilde{\boldsymbol{R}}^{*}=(0,0,2h)^{\text{tr}}.$
We define $\boldsymbol{\nabla}^{*}=\boldsymbol{\nabla}_{\boldsymbol{r}^{*}}$,
$\mathcal{M}_{\beta\gamma}^{f}=\frac{1-\lambda^{f}}{1+\lambda^{f}}\delta_{\beta\rho}\delta_{\rho\gamma}-\delta_{\beta z}\delta_{z\gamma}$,
the mirroring operator $\mathcal{M}_{\beta\gamma}=\delta_{\beta\rho}\delta_{\rho\gamma}-\delta_{\beta z}\delta_{z\gamma}$
and the index $\rho=x,y$ in the plane of the interface.}
\end{table*}
\begin{figure}
\centering 
\includegraphics[width=0.35\columnwidth]{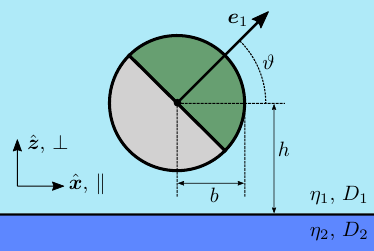}
\vspace{0mm}\caption{\label{fig:schematic}\textbf{Half-coated phoretic particle near the
surface of two immiscible liquids.} Schematic representation of a
half-coated phoretic particle (Janus particle) of radius $b$ and
with orientation $\bm{e}_{1}$ at a height $h$ above an interface
of viscosity ratio $\lambda^{f}=\eta_{2}/\eta_{1}$ and diffusivity
ratio $\lambda^{c}=D_{2}/D_{1}$. The latter effectively measures
how permeable the interface is to the solutes, where $\lambda^{c}=0$
implies an impermeable and $\lambda^{c}=1$ a perfectly permeable
interface. Due to the cylindrical symmetry of the system we can restrict
our attention to the $x$-$z$ plane, where the symbols $\parallel$
and $\perp$ imply motion parallel and perpendicular to the interface,
respectively. }
\end{figure}
We now introduce a plane surface of two immiscible liquids of viscosity
ratio $\lambda^{f}=\eta_{2}/\eta_{1}$ and solute diffusivity ratio
$\lambda^{c}=D_{2}/D_{1}$ in the vicinity of the particle, see Figure \ref{fig:schematic}. The interface
is characterised by the Green's functions in Table \ref{tab:GREEN}.
Here, $H$ satisfies the boundary condition of continuous normal flux
$\hat{\boldsymbol{z}}\cdot\boldsymbol{j}^{(1)}=\hat{\boldsymbol{z}}\cdot\boldsymbol{j}^{(2)}$
across the interface and $\boldsymbol{G}$ arises from the boundary conditions of continuous
tangential flow $v_{\rho}^{(1)}=v_{\rho}^{(2)},$ vanishing normal
flow $v_{z}^{(1)}=v_{z}^{(2)}=0$ and continuous tangential stress
$\eta_{1}\sigma_{\rho z}^{(1)}=\eta_{2}\sigma_{\rho z}^{(2)}$ across
the interface, where the index $\rho=x,y$ lies in the plane of the
interface \citep{jonesDiffusionPolymersFluidFluid1975,aderogbaActionForcePlanar1978}.
The superscripts label whether the quantity of interest is above or
below the interface, where $(1)$ refers to the positive half-space
$z>0$.

To discuss the chemo-hydrodynamic effect a plane interface has on
autophoresis and Brownian motion, we choose a simple non-axisymmetric
particle model. We truncate the expansions of the phoretic mobility
and surface flux each at linear order and choose $J_{0}/3=J_{1}=J$,
so that the particle has one inert pole ($j^{\mathcal{A}}=0$) and
one active pole ($j^{\mathcal{A}}>0$ ($j^{\mathcal{A}}<0$) for $J>0$
($J<0$), corresponding to a source (sink) of chemical reactants).
This is a first-order approximation to a Janus swimmer so that for
$J>0$ we have an inert-side-forward swimmer. We define $\bm{e}_{1}\cdot\bm{p}_{1}=\cos\alpha$,
where as before, $\bm{e}_{1}$ shall serve as the orientation of the
particle. For $\sin\alpha\neq0$ the particle is capable of phoretic
self-rotation, see the discussion on circular swimming in the previous
section.

We choose to truncate the generated concentration field at the surface
of the particle at second order with the coefficients determined by
Eq. (\ref{eq:conc-general}). Since slip is proportional to gradients
in the surface concentration we can ignore the terms $\boldsymbol{\zeta}^{(0,q)}$,
$\boldsymbol{\zeta}^{(q,0)}$ and $\boldsymbol{\mathcal{E}}^{(0,q)}$.
Cylindrical symmetry of the system can then be used to write the remaining
non-zero chemical tensors in terms of scalar coefficients, which are
given in Table \ref{tab:Elastance-coefficients}. For simplicity,
we assume the absence of any background concentration field.
\begin{table}
\begin{tabular}{>{\centering}p{2cm}>{\raggedright}p{9cm}}
\toprule 
\addlinespace
\textbf{Tensors} & \textbf{$\quad$Scalar coefficients}\tabularnewline\addlinespace
\midrule
\midrule 
\addlinespace
$\boldsymbol{\zeta}^{(1,1)}$ & $\quad\zeta_{\parallel}^{(1,1)}=\zeta_{1}\left[1+\frac{1}{16}\Lambda^{c}\hat{h}^{-3}\right]$,$\qquad$$\quad\zeta_{\perp}^{(1,1)}=\zeta_{1}\left[1+\frac{1}{8}\Lambda^{c}\hat{h}^{-3}\right]$\tabularnewline\addlinespace
\midrule 
\addlinespace
$\boldsymbol{\mathcal{E}}^{(1,0)}$ & $\quad\mathcal{E}^{(1,0)}=-\tfrac{1}{4}\mathcal{E}_{1}\Lambda^{c}\hat{h}^{-2}$\tabularnewline\addlinespace
\midrule 
\addlinespace
$\boldsymbol{\mathcal{E}}^{(1,1)}$ & $\quad\mathcal{E}_{\parallel}^{(1,1)}=\mathcal{E}_{1}\left[1+\frac{3}{16}\Lambda^{c}\hat{h}^{-3}\right]$,$\qquad$$\quad\mathcal{E}_{\perp}^{(1,1)}=\mathcal{E}_{1}\left[1+\frac{3}{8}\Lambda^{c}\hat{h}^{-3}\right]$\tabularnewline\addlinespace
\midrule 
\addlinespace
$\boldsymbol{\mathcal{E}}^{(2,0)}$ & $\quad\mathcal{E}^{(2,0)}=-\tfrac{1}{48}\mathcal{E}_{2}\Lambda^{c}\hat{h}^{-3}$\tabularnewline\addlinespace
\midrule 
\addlinespace
$\boldsymbol{\mathcal{E}}^{(2,1)}$ & $\quad\mathcal{E}^{(2,1)}=-\tfrac{3}{64}\mathcal{E}_{2}\Lambda^{c}\hat{h}^{-4}$\tabularnewline\addlinespace
\bottomrule
\end{tabular}\caption{\label{tab:Elastance-coefficients}\textbf{Chemical coefficients.}
Scalar coefficients for the linear response to a background concentration
field and for the elastance tensors with $\Lambda^{c}=\left(1-\lambda^{c}\right)/\left(1+\lambda^{c}\right)$,
where $\lambda^{c}=D_{2}/D_{1}$, and the height above the interface
$\hat{h}=h/b$. We have used the exact unbounded coefficients in Eq.
(\ref{eq:exact-chem-coeff}) such that $\zeta_{1}=3/2$, $\mathcal{E}_{1}=3/8\pi bD_{1},$
and $\mathcal{E}_{2}=5/2\pi bD_{1}$. Cylindrical symmetry of the
system implies for the coupling to a linear background concentration
field, $\boldsymbol{\zeta}^{(1,1)}=(\boldsymbol{I}-\hat{\boldsymbol{z}}\hat{\boldsymbol{z}})\,\zeta_{\parallel}^{(1,1)}+\hat{\boldsymbol{z}}\hat{\boldsymbol{z}}\,\zeta_{\perp}^{(1,1)}$.
The relevant elastance tensors are $\boldsymbol{\mathcal{E}}^{(1,0)}=\hat{\boldsymbol{z}}\,\mathcal{E}^{(1,0)},$
$\boldsymbol{\mathcal{E}}^{(1,1)}=(\boldsymbol{I}-\hat{\boldsymbol{z}}\hat{\boldsymbol{z}})\,\mathcal{E}_{\parallel}^{(1,1)}+\hat{\boldsymbol{z}}\hat{\boldsymbol{z}}\,\mathcal{E}_{\perp}^{(1,1)},$
$\boldsymbol{\mathcal{E}}^{(2,0)}=-\left(3\hat{\boldsymbol{z}}\hat{\boldsymbol{z}}-\boldsymbol{I}\right)\,\mathcal{E}^{(2,0)},$
and $\mathcal{E}_{\alpha\beta\gamma}^{(2,1)}=\mathcal{E}^{(2,1)}\left[\left(\delta_{\gamma\alpha}-\delta_{\gamma z}\delta_{\alpha z}\right)\delta_{\beta z}+\left(\delta_{\gamma\beta}-\delta_{\gamma z}\delta_{\beta z}\right)\delta_{\alpha z}+\left(3\delta_{\beta z}\delta_{\alpha z}-\delta_{\alpha\beta}\right)\delta_{\gamma z}\right].$}
\end{table}

The induced slip sets the surrounding fluid in motion. The fluid then
reacts back on the particle and causes rigid-body motion, governed
by the equations of motion in Eq. (\ref{eq:EoM}), mediated by mobility
and propulsion tensors. Again, the cylindrical symmetry of the system
allows us to write the mobility and propulsion tensors in terms of
scalar coefficients, summarised in Table \ref{tab:Mobility-coefficients}.
\begin{table}
\begin{tabular}{>{\centering}p{2cm}>{\raggedright}p{9.5cm}}
\toprule 
\textbf{Tensors} & \textbf{$\quad$Scalar coefficients}\tabularnewline
\midrule
\midrule 
\addlinespace
$\boldsymbol{\mu}^{TT}$ & $\quad\mu_{\parallel}^{TT}=\mu_{T}\left[1+\frac{3(2-3\lambda^{f})}{16(1+\lambda^{f})}\hat{h}^{-1}+\frac{1+2\lambda^{f}}{16(1+\lambda^{f})}\hat{h}^{-3}-\frac{\lambda^{f}}{16(1+\lambda^{f})}\hat{h}^{-5}\right]$\tabularnewline\addlinespace
\addlinespace
 & $\quad\mu_{\perp}^{TT}=\mu_{T}\left[1-\frac{3(2+3\lambda^{f})}{8(1+\lambda^{f})}\hat{h}^{-1}+\frac{1+4\lambda^{f}}{8(1+\lambda^{f})}\hat{h}^{-3}-\frac{\lambda^{f}}{8(1+\lambda^{f})}\hat{h}^{-5}\right]$\tabularnewline\addlinespace
\midrule 
\addlinespace
$\boldsymbol{\mu}^{RR}$ & $\quad\mu_{\parallel}^{RR}=\mu_{R}\left[1+\frac{1-5\lambda^{f}}{16(1+\lambda^{f})}\hat{h}^{-3}\right]$\tabularnewline\addlinespace
\addlinespace
 & $\quad\mu_{\perp}^{RR}=\mu_{R}\left[1+\frac{1-\lambda^{f}}{8(1+\lambda^{f})}\hat{h}^{-3}\right]$\tabularnewline\addlinespace
\midrule 
\addlinespace
$\boldsymbol{\mu}^{TR}$ & $\quad\mu^{TR}=\frac{1}{6\pi\eta b^{2}}\left[-\frac{3}{16(1+\lambda^{f})}\hat{h}^{-2}+\frac{3\lambda^{f}}{32(1+\lambda^{f})}\hat{h}^{-4}\right]$\tabularnewline\addlinespace
\midrule 
\addlinespace
$\boldsymbol{\pi}^{(T,2s)}$ & $\quad\pi_{1}^{(T,2s)}=\frac{5\lambda^{f}}{16(1+\lambda^{f})}\hat{h}^{-2}-\frac{1+3\lambda^{f}}{12(1+\lambda^{f})}\hat{h}^{-4}+\frac{5\lambda^{f}}{48(1+\lambda^{f})}\hat{h}^{-6}$\tabularnewline\addlinespace
\addlinespace
 & $\quad\pi_{2}^{(T,2s)}=-\frac{5\left(2+3\lambda^{f}\right)}{48(1+\lambda^{f})}\hat{h}^{-2}+\frac{4+15\lambda^{f}}{48(1+\lambda^{f})}\hat{h}^{-4}-\frac{5\lambda^{f}}{48(1+\lambda^{f})}\hat{h}^{-6}$\tabularnewline\addlinespace
\midrule 
\addlinespace
$\boldsymbol{\pi}^{(T,3t)}$ & $\quad\pi_{\parallel}^{(T,3t)}=-\frac{1+2\lambda^{f}}{80(1+\lambda^{f})}\hat{h}^{-3}+\frac{\lambda^{f}}{40(1+\lambda^{f})}\hat{h}^{-5}$\tabularnewline\addlinespace
\addlinespace
 & $\quad\pi_{\perp}^{(T,3t)}=-\frac{1+4\lambda^{f}}{40(1+\lambda^{f})}\hat{h}^{-3}+\frac{\lambda^{f}}{20(1+\lambda^{f})}\hat{h}^{-5}$\tabularnewline\addlinespace
\midrule 
\addlinespace
$\boldsymbol{\pi}^{(T,4t)}$ & $\quad\pi_{1}^{(T,4t)}=\frac{1+3\lambda^{f}}{672\left(1+\lambda^{f}\right)}\hat{h}^{-4}-\frac{5\lambda^{f}}{1008\left(1+\lambda^{f}\right)}\hat{h}^{-6}$\tabularnewline\addlinespace
\addlinespace
 & $\quad\pi_{2}^{(T,4t)}=-\frac{1+5\lambda^{f}}{672\left(1+\lambda^{f}\right)}\hat{h}^{-4}+\frac{5\lambda^{f}}{1008\left(1+\lambda^{f}\right)}\hat{h}^{-6}$\tabularnewline\addlinespace
\midrule 
\addlinespace
$\boldsymbol{\pi}^{(R,2s)}$ & $\quad\pi^{(R,2s)}=\frac{1}{b}\left[\frac{5}{32}\hat{h}^{-3}-\frac{\lambda^{f}}{8(1+\lambda^{f})}\hat{h}^{-5}\right]$\tabularnewline\addlinespace
\midrule 
\addlinespace
$\boldsymbol{\pi}^{(R,3t)}$ & $\quad\pi^{(R,3t)}=\frac{1}{b}\frac{3\lambda^{f}}{80(1+\lambda^{f})}\hat{h}^{-4}$\tabularnewline\addlinespace
\midrule 
\addlinespace
$\boldsymbol{\pi}^{(R,4t)}$ & $\quad\pi^{(R,4t)}=\frac{1}{b}\frac{\lambda^{f}}{168\left(1+\lambda^{f}\right)}\hat{h}^{-5}$\tabularnewline\addlinespace
\bottomrule
\end{tabular}

\caption{\label{tab:Mobility-coefficients}\textbf{Hydrodynamic coefficients.}
Scalar coefficients for the mobility matrices and the relevant propulsion
tensors with $\lambda^{f}=\eta_{2}/\eta_{1}$ and the height above
the interface $\hat{h}=h/b$. Cylindrical symmetry of the system allows
us to write for the translational mobilities, $\boldsymbol{\mu}^{TT}=(\boldsymbol{I}-\hat{\boldsymbol{z}}\hat{\boldsymbol{z}})\mu_{\parallel}^{TT}+\hat{\boldsymbol{z}}\hat{\boldsymbol{z}}\mu_{\perp}^{TT}$
and $\boldsymbol{\mu}^{TR}=\left(\boldsymbol{\mu}^{RT}\right)^{\text{tr}}=\mu^{TR}\boldsymbol{\epsilon}\cdot\hat{\boldsymbol{z}}$.
The mobility $\boldsymbol{\mu}^{RR}$ has the same structure as $\boldsymbol{\mu}^{TT}$
with the corresponding coefficients $\mu_{\parallel}^{RR}$ and $\mu_{\perp}^{RR}$.
The propulsion tensor for the leading symmetric slip mode is $\ensuremath{\pi_{\alpha\beta\gamma}^{(T,2s)}}=\pi_{1}^{(T,2s)}\left[\left(\delta_{\gamma\alpha}-\delta_{\gamma z}\delta_{\alpha z}\right)\delta_{\beta z}+\left(\delta_{\beta\alpha}-\delta_{\beta z}\delta_{\alpha z}\right)\delta_{\gamma z}\right]+\pi_{2}^{(T,2s)}\left(\delta_{\gamma\beta}-3\delta_{\gamma z}\delta_{\beta z}\right)\delta_{\alpha z}$.
The propulsion tensor $\boldsymbol{\pi}^{(T,4t)}$ is of the same
structure as $\boldsymbol{\pi}^{(T,2s)}$ with the corresponding coefficients
$\pi_{1}^{(T,4t)}$ and $\pi_{2}^{(T,4t)}$. Since $\boldsymbol{\pi}^{(T,3t)}$
has the same structure as $\boldsymbol{\mu}^{TT}$, we adopt an analogous
notation with $\pi_{\parallel}^{(T,3t)}$ and $\pi_{\perp}^{(T,3t)}$.
For the propulsion tensors contributing to the particle's rotational
dynamics we obtain $\pi_{\alpha\beta\gamma}^{(R,2s)}=\pi^{(R,2s)}\left(\delta_{\beta z}\epsilon_{z\gamma\alpha}+\delta_{\gamma z}\epsilon_{z\beta\alpha}\right)$
and $\boldsymbol{\pi}^{(R,3t)}=\pi^{(R,3t)}\boldsymbol{\epsilon}\cdot\hat{\boldsymbol{z}}$.
The propulsion tensor $\boldsymbol{\pi}^{(R,4t)}$ is of the same
structure as $\boldsymbol{\pi}^{(R,2s)}$ with the corresponding coefficient
$\pi^{(R,4t)}$.}
\end{table}

The full set of mobility coefficients for a wall, a free surface or,
indeed, a fluid-fluid interface have been obtained in the literature
to a high degree of accuracy \citep{BRENNER1961242,goldmanSlowViscousMotion,felderhofForceDensityInduced1976,leeMotionSpherePresence1979,leeMotionSpherePresence1980,perkinsHydrodynamicInteractionSpherical1990,perkinsHydrodynamicInteractionSpherical1992}.
In Appendix \ref{sec:Simulation} we show that for the plane boundary
the diffusion terms $\propto\sqrt{2k_{B}T\,\mathbb{M}}$ in Eqs. (\ref{eq:EoM})
take a particularly simple analytic form and that with the coefficients
given in Table \ref{tab:Mobility-coefficients} this diffusion matrix
is inherently positive-definite for all physical configurations. The
propulsion tensors are a unique feature of active particles and have
not been obtained in this form in the literature before. 

\begin{figure}
\centering 
\includegraphics[width=\columnwidth]{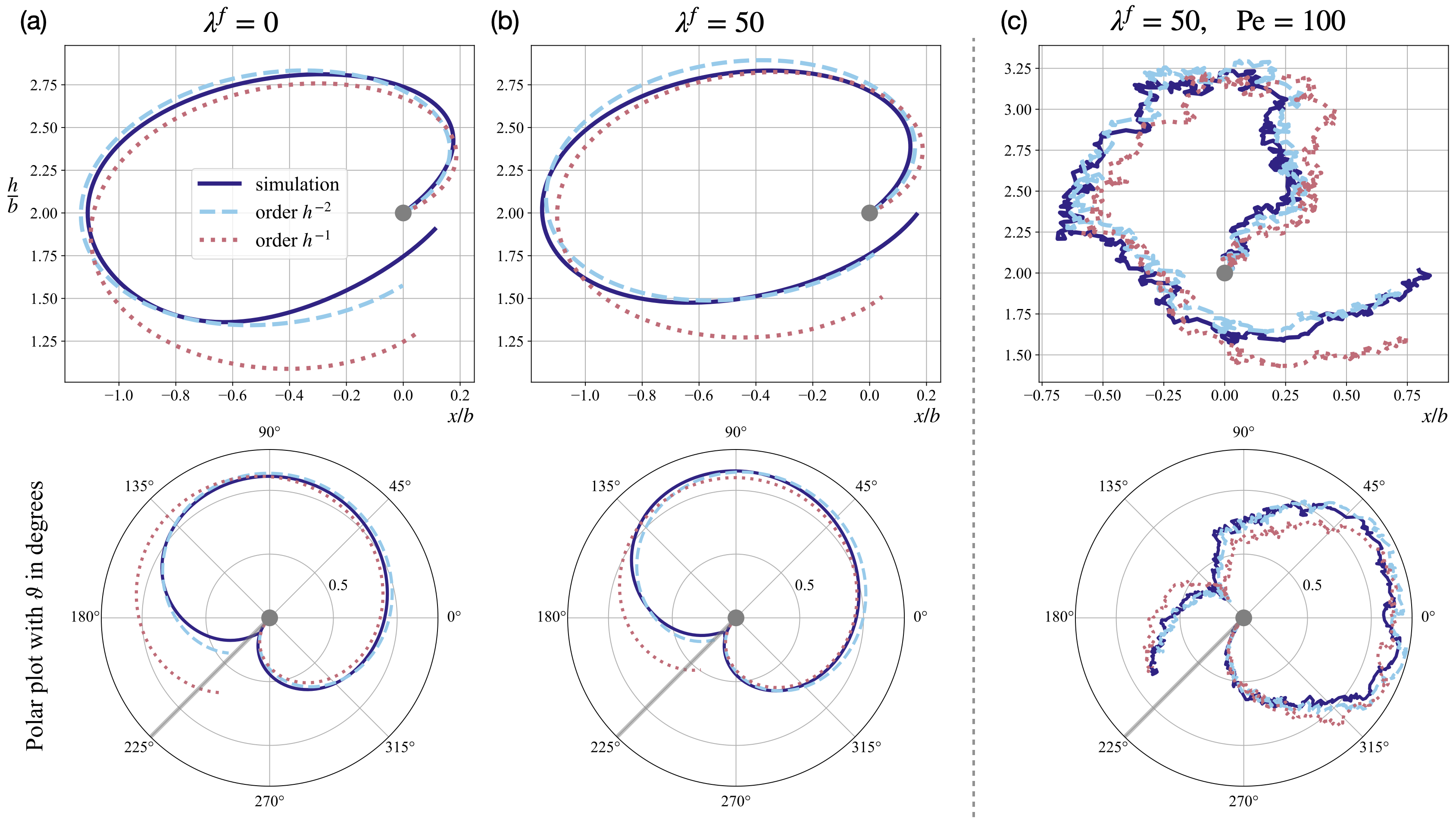}
\vspace{-3mm}\caption{\label{fig:simulation}\textbf{Chemo-hydrodynamic effects of a boundary.}
We compare typical trajectories of a bottom-heavy non-axisymmetric
active particle (see the main text for the chosen particle specifications)
near two types of chemically impermeable ($\lambda^{c}=0$) liquid-liquid
interfaces. In panels (a) and (b) the interface is chosen to be a
free air-water surface ($\lambda^{f}=0$) and a water-oil interface
($\lambda^{f}\approx50$), respectively. The temperature in these
panels is zero and the starting point (orientation) of the particle
is indicated by a grey disk (line). In the upper panels we compare
the trajectories of the particle obtained to various orders in the
inverse distance to the wall $h^{-1}$, thus gradually including more
interactions with the wall. Here, `simulation' refers to the full
dynamical system at the accuracy obtained in this paper. The lower
panels show the orientational evolution of the particle in the various
approximations in a polar plot, parametrised by the angle $\vartheta$
and the distance to the starting point. There are two main features
to be observed when comparing (a) and (b). First, the particle tends
to stay further away from the water-oil interface when compared with
the air-water surface. This is expected as the particle's mobility
perpendicular to the interface will be reduced with increasing viscosity
ratio, see Table \ref{tab:Mobility-coefficients}. Second, the approximations
are better for higher viscosity ratios. Again, this can be understood
intuitively when considering the example that fluid flows produced
near a wall decay faster in the far-field when compared with fluid flows
produced near a free surface \citep{aderogbaActionForcePlanar1978}.
The slower decay of the latter means that to achieve the same quality
of approximation as with interfaces of higher viscosity ratios, higher
orders in $h^{-1}$ are necessary. Panel (c) shows the same system
as in panel (b) but with a finite particle P\'{e}clet number of ${\rm Pe}=100$,
representing experimentally relevant noise levels. It is clear that
the translational as well as the rotational diffusion is affected
by the presence of the interface. Thermal diffusion also induces a
net repulsive effect between the particle and the interface. }
\end{figure}
Assuming there is no external torque rotating the particle out of
plane, i.e., $\boldsymbol{T}^{P}\cdot\hat{\boldsymbol{z}}=0$, we
can once again restrict our attention to the $x$-$z$ plane for which
we define the planar polar angle $\vartheta$ such that $\boldsymbol{e}_{1}=\cos\vartheta\,\hat{\boldsymbol{x}}+\sin\vartheta\,\hat{\boldsymbol{z}}.$
Autophoretic particles in typical experiments are neither force nor
torque free due to mismatches between particle and solvent densities
and between gravitational and geometric centres \citep{ebbens2010pursuit,drescherDirectMeasurementFlow2010,palacci2010sedimentation,palacci2013living,buttinoni2013DynamicClustering}.
Since the resulting forces and torques become dominant, at long distances,
over active contributions, it is crucial to include their effects
in our analysis. In simulating Eq. (\ref{eq:dynamics}) we therefore
assume a bottom-heavy Janus particle (the chemically active coating,
blue in figures, is assumed to be slightly heavier than the inert
side, white in figures). Therefore, we have to take into account gravity
in negative $z$-direction and a gravitational torque given by
\begin{equation}
\boldsymbol{F}^{P}=-mg\hat{\boldsymbol{z}},\qquad\boldsymbol{T}^{P}=\kappa(\hat{\boldsymbol{z}}\times\boldsymbol{e})=\kappa\cos\vartheta\,\hat{\boldsymbol{y}},
\end{equation}
with $m$ the buoyancy-corrected mass of the particle, $g$ the gravitational
constant and $\kappa$ a constant parametrising bottom-heaviness.
Inserting this into the update equations (\ref{eq:EoM}) with $\boldsymbol{R}=(x,y,h)^{\text{tr}}$,
where $h(t)$ is the height of the particle above the boundary, we
can now simulate the time evolution of this bottom-heavy, non-axisymmetric
phoretic particle. In Figure \ref{fig:simulation} we show typical
trajectories near a free surface (e.g. an air-water surface) and a
fluid-fluid interface of $\lambda^{f}=50$ (e.g. an oil-water interface).
We probe the effect of the nearby boundary on the dynamics of the
autophoretic particle by truncating the dynamical system in Eq. (\ref{eq:dynamics})
at various orders in $h^{-1}$ and comparing the results, see Appendix
\ref{sec:Simulation} for the truncated expressions.

At order $h^{-1}$ only hydrodynamic interactions with the boundary
due to the gravitational force manifest themselves. It is at the next order, $h^{-2}$, that the gravitational torque
 and active effects become apparent. The latter comprise hydrodynamic
interactions from symmetric propulsion via $\boldsymbol{\pi}^{(T,2s)}$
and a purely monopolar chemical interaction with the interface. At
this order in the approximation, fore-aft symmetry breaking of the
particle is no longer necessary for self-propulsion near a boundary,
see Appendix \ref{sec:Simulation}. An isotropic particle with uniform
phoretic mobility $\mu_{c}$ and surface flux $j^{\mathcal{A}}$ will
get repelled (attracted) to the interface depending on whether it
is a source or sink of chemical reactants and depending on the diffusivity
ratio $\lambda^{c}$ of the interface, see Section \ref{subsec:hovering}
for a detailed discussion. Furthermore, at this order in the approximation
the thermal advective term in Eq. (\ref{eq:EoM}) proportional to
$\boldsymbol{\nabla}\cdot\mathbb{M}$ starts to affect the dynamics.
It is worth noting that, at order $h^{-3}$, our analytical results
match those obtained in Ref. \citep{ibrahimDynamicsSelfphoreticJanus2015},
using a method of reflections, for a Janus particle of trivial phoretic
mobility near an inert no-slip wall.

The system parameters in Figure \ref{fig:simulation} are chosen as
follows. The starting position of the particle is at a height $\hat{h}_{0}=h(t=0)/b=2$
and an angle $\vartheta_{0}=-3\pi/4$ to the wall. For the surface
flux of the particle we choose the dimensionless control parameter
$j_{1}=J_{1}/J_{0}=1/3$, modelling a source particle. Its phoretic
mobility distribution is specified by the dimensionless parameter
$m_{1}=M_{1}/M_{0}=0.7$, implying a significant non-isotropy ($m_{1}=0$
specifies a trivial phoretic mobility). The angle between the axes
of surface flux and phoretic mobility is chosen such that $\alpha=\pi/2$.
In Figure \ref{fig:simulation}(c) the Brown number is set to $\mathcal{B}\sim10^{-2}$,
roughly matching a set of experiments on Janus colloids \citep{jiang2010active,palacci2013living}
with a bead size $b\sim1\mu m$ and speed $v_{s}\sim10\mu ms^{-1}$.
The ratios of gravitational to active forces and torques are chosen
such that $mg/F_{A}\sim10^{-1}$ and $\kappa/T_{A}\sim10^{-2}$, respectively.
Finally, inertial effects decay on the time scale of momentum relaxation,
typically $\tau_{T}=m/6\pi\eta b$ and $\tau_{R}=I/8\pi\eta b^{3}=m/20\pi\eta b$
for translational and rotational effects, respectively. The time step
$\Delta t$ in our simulation is chosen such that $\tau_{T}/\Delta t\sim10^{-4}$
and $\tau_{R}/\Delta t\sim10^{-4}$, ensuring that the Smoluchowski
limit of the dynamics provides an appropriate description.

\subsection{Hovering above a permeable interface\label{subsec:hovering}}

As mentioned in the previous section, if chemo-hydrodynamic particle-boundary
interactions of order $h^{-2}$ and higher are considered, fore-aft
symmetry breaking of the particle is no longer necessary for self-propulsion
near a boundary. Indeed, self-propulsion of isotropic particles near
a boundary has been observed in light-activated phoretic swimmers
\citep{palacci2013living}. We therefore consider a particle that
is an isotropic source of reactants ($\mu_{c}={\rm const}>0$, $j^{\mathcal{A}}={\rm const}>0$)
and investigate how its dynamics are affected by the viscosity ratio
$\lambda^{f}$ and the diffusivity ratio $\lambda^c$ of the nearby interface in the limit of zero temperature.
The particle is assumed to be driven towards the interface due to
gravity. With the rotational dynamics and the translational dynamics
parallel to the plane vanishing by symmetry, the particle is expected
to hover above the interface at a height that can be found by solving
$\dot{h}=0$, where $\dot{h}$ is given in Appendix \ref{sec:Simulation}.
Rescaling heights by $b$, mobilities by $\mu_{T}$ and velocities
by $\mu_{T}mg$ and renaming the thus non-dimensionalised variables
such that they read the same, we obtain the hovering condition
\begin{equation}
0=-\mu_{\perp}^{TT}+\tfrac{1}{4}\Lambda^{c}\mathcal{A}_{G}\left[h^{-2}(1+5\pi_{\perp}^{(T,3t)})-3h^{-3}(\pi_{2}^{(T,2s)}-14\pi_{2}^{(T,4t)})\right].\label{eq:hovering}
\end{equation}
Hovering is thus characterised by only one dimensionless number, $\mathcal{A}_{G}=\mu_{c}j^{\mathcal{A}}/D\mu_{T}mg$,
defined as the ratio of the speed of a uniformly coated phoretic particle
in a uniform concentration gradient $j^{\mathcal{A}}/D$, namely $\mu_{c}j^{\mathcal{A}}/D$,
to the settling velocity under gravity $\mu_{T}mg$. This is a measure
of activity.

\begin{figure}
\centering 
\includegraphics[width=0.9\columnwidth]{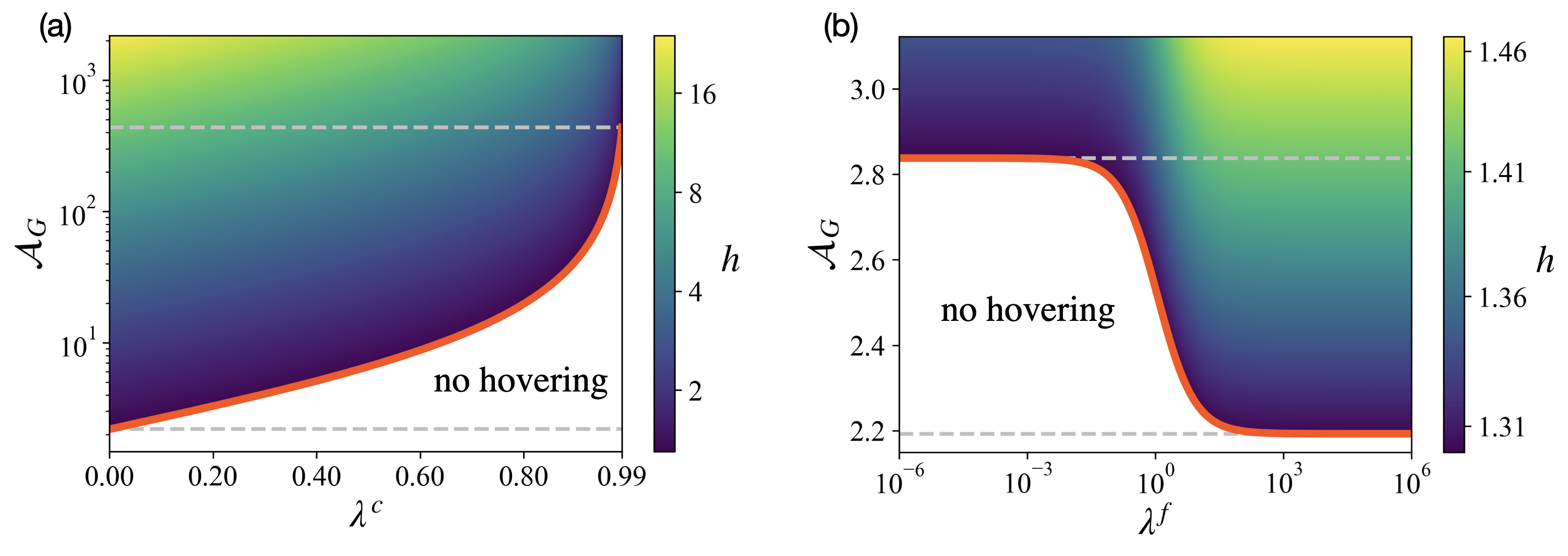}
\vspace{-3mm}\caption{\label{fig:hovering}\textbf{Hovering above an interface.} We show
the hovering height $h$ (pseudo-colour map) for an active particle
at zero temperature as a function of its activity $\mathcal{A}_{G}$
and the diffusivity ratio $\lambda^{c}$ or the viscosity ratio $\lambda^{f}$
of the interface. In panel (a) we consider the particle hovering above
a wall ($\lambda^{f}\rightarrow\infty$) that is permeable to the
solutes. For a boundary between regions of equal solute diffusivity ($\lambda^{c}=1$) there
exists no chemical repulsion, leading to the particle inevitably crashing
into the boundary. Therefore, we only consider values $\lambda^{c}\protect\leq0.99$.
The red line shows the particle's minimum activity to hover at a minimum
height of $h_{{\rm min}}=1.3$ as a function of the diffusivity ratio.
The white region below the red line indicates physics that is not
accessible in our simplified model and we assume that the particle
crashes into the boundary due to gravity, i.e., we set $h=1$. The
two dashed grey lines indicate the limiting values $\mathcal{A}_{G}^{1}\approx2.2$
and $\mathcal{A}_{G}^{2}\approx440$ that are required to hover above
an impermeable and a highly permeable wall. In panel (b) we consider
the particle hovering above an impermeable ($\lambda^{c}=0$) interface
of varying viscosity ratio. The two horizontal dashed grey lines indicate
the limiting values $\mathcal{A}_{G}^{3}\approx2.8$ and $\mathcal{A}_{G}^{1}$
that are required to hover above a free surface and a solid wall,
respectively.}
\end{figure}
\begin{figure}
\centering 
\includegraphics[width=\columnwidth]{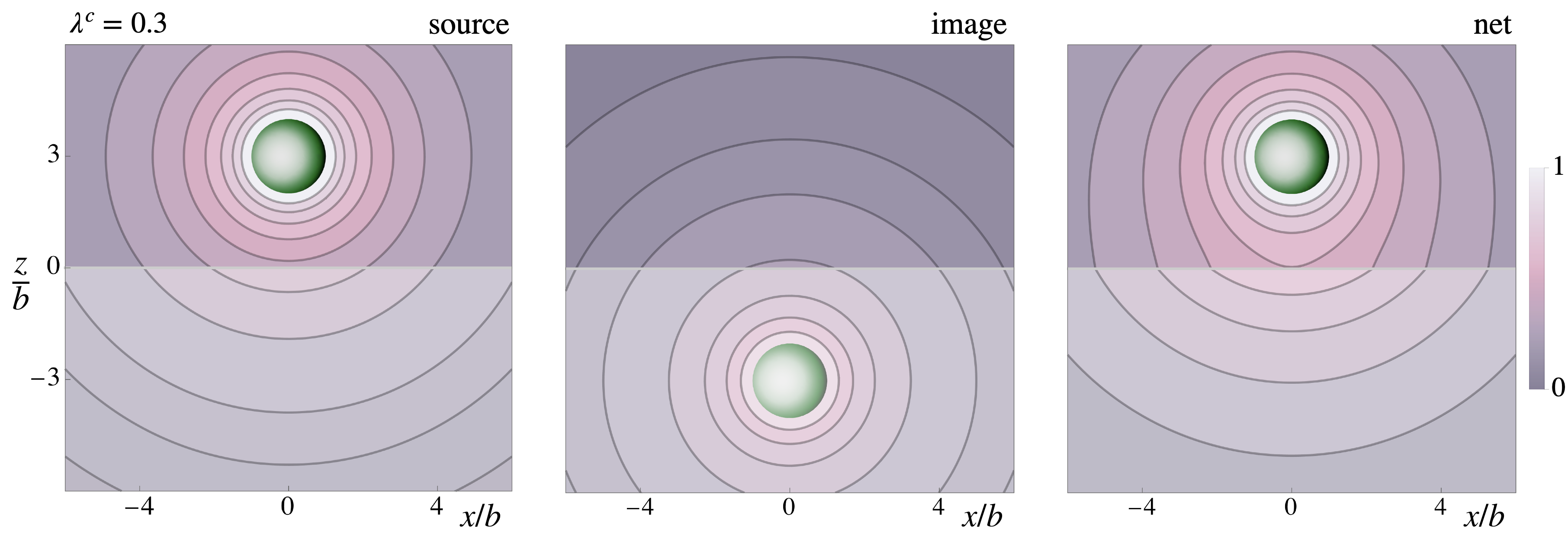}
\vspace{-3mm}\caption{\label{fig:conc-hover}\textbf{Solute concentration for a particle
hovering above a permeable boundary.} The chemical field generated
by a hovering source particle (green) is shown above ($z>0$) and
below ($z<0$) the permeable ($\lambda^{c}=0.3$) boundary as a pseudo-colour
map, where contours indicate regions of constant concentration. We
emphasise that the net concentration (right panel) in the region containing
the particle is a superposition of the source without the boundary
present (left panel) and its image below the boundary (middle panel).
The net concentration below the permeable boundary is then generated
by the appropriate boundary conditions. To imply a change in the chemical
diffusivity $D_{2}=\lambda^{c}D_{1}$ the colour map for the region
$z<0$ is shown with slightly reduced opacity. Since source particles
want to move down chemical gradients (anti-chemotaxis), it is clear
that the image creates a repulsive concentration field in $z>0$,
making it possible for the particle to hover above the boundary. It
is worth noting that, for an impermeable boundary ($\lambda^{c}=0$),
the contour lines meet the boundary at a right angle and the corresponding
vector field ($\bm{\nabla}c$) becomes purely tangential to this `no-flux'
boundary. In the limit of rapid solute diffusion the viscosity ratio
of the interface has no influence on the chemical field. }
\end{figure}
\begin{figure}
\centering 
\includegraphics[width=\columnwidth]{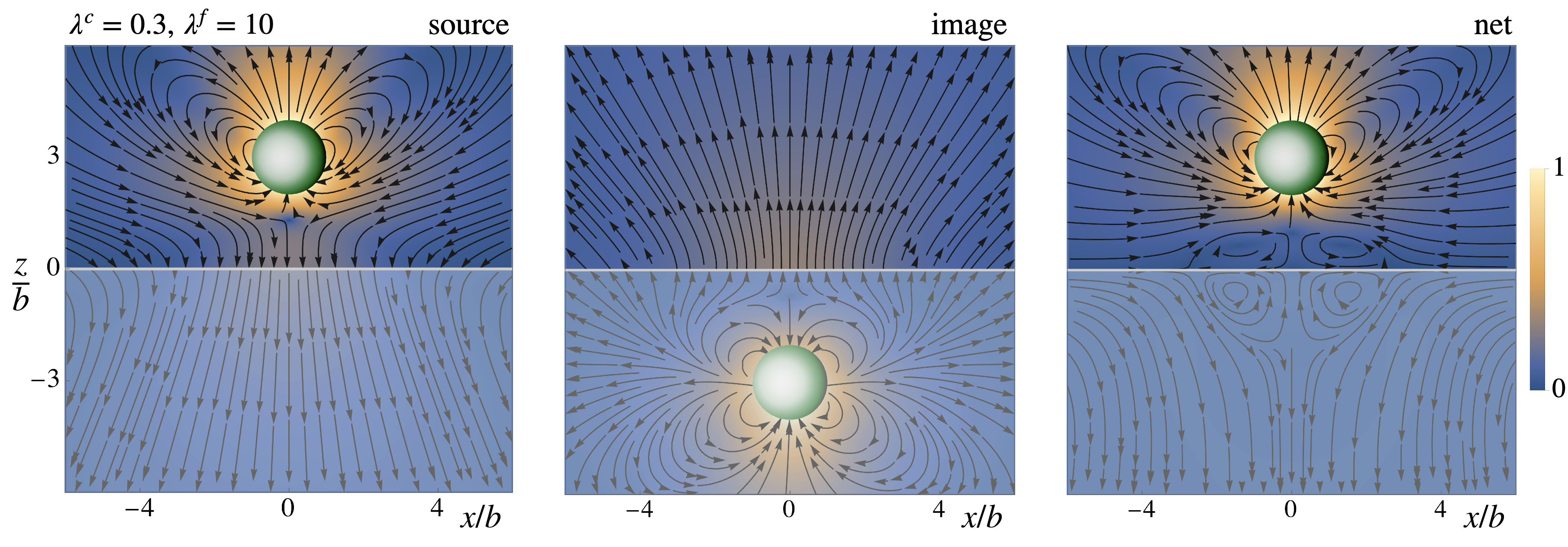}
\vspace{-3mm}\caption{\label{fig:flow-hover}\textbf{Fluid flow for a particle hovering
above a fluid-fluid interface.} The fluid flow (laboratory frame) generated
by a hovering source particle (green) is shown above ($z>0$) and
below ($z<0$) a chemically permeable fluid-fluid interface ($\lambda^{c}=0.3$,
$\lambda^{f}=10$). The direction of the fluid flow is indicated by
black arrows, while its relative magnitude is implied by the overlaid
pseudo-colour map. The net flow (right panel) in the region containing
the particle is a superposition of the source without the boundary
conditions on the fluid flow and stress (left panel) and its image
below the boundary (middle panel). Note that for the source the chemical
boundary conditions must still be satisfied (see the net chemical
field in Figure \ref{fig:conc-hover}), inducing a non-trivial slip
on the otherwise isotropic particle. The net flow below the interface
is then driven by the appropriate velocity and stress boundary conditions.
To imply a change in the viscosity $\eta_{2}=\lambda^{f}\eta_{1}$
the colour map for the region $z<0$ is shown with slightly reduced
opacity. It is clear that the image produces an upwards flow in $z>0$
which balances gravity and makes the particle hover away from the
interface. In the net flow this creates convection rolls between the
particle and the interface which in turn drive convection rolls below
the interface. }
\end{figure}

In Figure \ref{fig:hovering}, we show how in our approximation the
hovering height $h$ of the isotropic particle is affected by its
activity, the diffusivity ratio and the viscosity ratio of the boundary.
We limit our results to $h>h_{\text{min}}=1.3$. This is because very
close to the interface other effects such as electric double-layer
and Van der Waals interactions are expected to dominate \citep{wuDirectMeasurementSingle2005,verweijHeightDistributionOrientation2020}.
As expected, Figure \ref{fig:hovering}(a) shows that a higher chemical
diffusivity ratio of the interface, and thus decreased chemical repulsion
from it, requires higher particle activities for hovering to remain
possible. From Figure \ref{fig:hovering}(b) it is clear that lower
levels of activity are sufficient for hovering above a solid wall
as compared with a free surface. This is intuitive when considering
that due to increased fluid internal friction (decreased mobility)
gravity becomes less effective in moving the particle towards a solid
wall compared to moving it towards a free surface (see the coefficient
$\mu_{\perp}^{TT}$ in Table \ref{tab:Mobility-coefficients}). Using
a method of images, in Figures \ref{fig:conc-hover} and \ref{fig:flow-hover}
we illustrate the chemical and hydrodynamic repulsion from a permeable
interface, respectively.

\section{Discussion\label{sec:Summary-and-outlook}}

In this paper, we have presented a simultaneous solution of the boundary-domain
integral equations describing the chemical and the fluid flow around
an autophoretic particle in a fluctuating environment. This has been
achieved in a basis of tensor spherical harmonics. Compared with the
common squirmer model approach to active particles \citep{lighthillSquirmingMotionNearly1952,blakeSphericalEnvelopeApproach1971,pakGeneralizedSquirmingMotion2014,pedleySquirmersSwirlModel2016},
our boundary-domain integral method offers the distinct advantage
of obtaining the traction on the particle directly in a complete orthonormal
basis. This provides a naturally kinetic approach via Newton's equations
in which thermal fluctuations manifest themselves as fluctuating stresses.
The Brownian motion of an autophoretic particle is obtained in terms
of coupled roto-translational stochastic update equations containing
mobility and propulsion tensors. The latter are found to arise from
chemical activity of the particle and the chemo-hydrodynamic coupling
at the particle's surface, inducing a coupling of slip modes. We have
obtained exact and leading-order solutions for both the chemical and
the fluctuating hydrodynamic problems far away from and in the vicinity
of boundaries, respectively. Studying the Brownian motion of particles
in the bulk, some of the flexibility of our method in particle design
has been demonstrated. In the case of autophoresis near a plane interface,
characterised by its solute diffusivity and viscosity ratios, we have
provided analytical expressions for the chemo-hydrodynamic coupling
tensors. The given mobilities ensure a positive-definite diffusion
matrix in stochastic simulations. Finally, we have studied the hovering
state of an isotropic phoretic particle above an interface as a function
of particle activity, and the diffusivity and viscosity ratios of
the interface. In doing so, we have provided numerical results as
well as physical insights into the repulsive chemo-hydrodynamic particle-interface
interactions.

We have given the leading-order results for the chemical and hydrodynamic
coupling tensors. In principle, these can be obtained to arbitrary
accuracy, and the general iterative solutions are given in the Appendix.
This non-trivial computation will be the topic of future work. While
our results in Section \ref{subsec:Slip-and-resulting} are guaranteed
to provide dissipative motion for physical configurations, in Brownian
simulations, unphysical situations with a non-zero particle-boundary
overlap may occur on occasion (the probability of which can be lowered
by imposing a short-range repulsive potential between the particle
and the boundary). In this case, one can either impose an \textit{ad hoc} regularisation
on the mobility \citep{wajnrybGeneralizationRotnePrager2013,singhFluctuatingHydrodynamicsBrownian2017,balboausabiagaBrownianDynamicsConfined2017}
or use a bounce-back condition, effectively implementing a reflective
boundary condition in the simulation \citep{volpeSimulationActiveBrownian2014}.

It is helpful to compare our results with previous work on chemical
and hydrodynamic interactions of an active particle in a fluctuating
fluid. We have shown that simultaneous harmonic expansions of the
surface fields provide a high degree of flexibility in particle design,
comparable to previous models capable of motion in three dimensions
\citep{lisickiAutophoreticMotionThree2018}. Additionally, our framework
has been shown to provide a straightforward way of studying the Brownian
dynamics of particles that, even in the limit of zero temperature, perform
complex motion \citep{vanteeffelenDynamicsBrownianCircle2008,mozaffariSelfpropelledColloidalParticle2018,baileyMinimalNumericalIngredients2023}.
To the best of our knowledge, this is the first work to obtain the
elastance for an active particle near an interface separating two
fluids of arbitrary diffusivity ratios. The corresponding Green's
function, which is given in Table \ref{tab:GREEN}, does not appear
anywhere in the literature, although its derivation is straightforward
given the correct boundary conditions. This paper is also the first
to simultaneously study the chemo-hydrodynamics of an autophoretic
particle near a planar surface of two immiscible fluids of arbitrary
ratio of viscosities and diffusivities. While previous works have
studied phoretic particles hovering above a chemically impermeable
solid wall as a function of particle coverage \citep{uspalSelfpropulsionCatalyticallyActive2015,ibrahimHowWallsAffect2016},
we investigated the hovering state as a function of the properties
of the interface, relevant for studies on particle aggregation near
fluid-fluid or free surfaces \citep{chen2015dynamic,hokmabadSpontaneouslyRotatingClusters2022}.

We briefly discuss the level of approximation of explicit results
provided in this paper. For a passive particle, the mobility of sedimentation
towards a plane interface is known exactly \citep{BRENNER1961242}.
In the absence of an exact solution for motion parallel to a boundary,
careful examination of the case when the particle-interface gap distance
is much smaller than the particle radius is necessary. In 1967, Goldman,
Cox and Brenner \citep{goldmanSlowViscousMotion1967} used a lubrication
approximation to derive an asymptotic solution for this case. However,
matching the asymptotic solutions for the near- ($h/b\ll1$) and far-field
($h/b\gg1$) limits can be challenging in dynamic simulations \citep{bradyStokesianDynamics,ichikiImprovementStokesianDynamics2002}.
It has been confirmed experimentally that for parallel motion the
order to which the mobilities are given in this paper provides a good
approximation even when the particle-interface gap distance is only
a fraction of the particle radius \citep{choudhuryActiveColloidalPropulsion2017}.
So while an approach using lubrication theory is appropriate for general
motion very close to a plane \citep{villaMotionMicroNano2020,villaMicroparticleBrownianMotion2023},
the given approximation in the mobilities arising from a series expansion
can still be expected to be of interest to a wide range of experimental
settings in which colloidal particles are studied near a plane boundary.
Thus, our work also adds to the existing literature on the mobility
\citep{BRENNER1961242,goldmanSlowViscousMotion,felderhofForceDensityInduced1976,perkinsHydrodynamicInteractionSpherical1990,perkinsHydrodynamicInteractionSpherical1992,leeMotionSpherePresence1979,leeMotionSpherePresence1980,swanSimulationHydrodynamicallyInteracting2007,daddi-moussa-iderSwimmingTrajectoriesThreesphere2018,michailidouDynamicsConcentratedHardSphere2009}
and diffusion \citep{ermakBrownianDynamicsHydrodynamic1978,wajnrybBrownianDynamicsDivergence2004,rogersRotationalDiffusionSpherical2012,delongBrownianDynamicsConfined2015,lisickiNearwallDiffusionTensor2016}
of a sedimenting particle near a boundary. Explicit expressions for
propulsion tensors and mobility matrices are given in Table \ref{tab:Mobility-coefficients},
while Table \ref{tab:Elastance-coefficients} contains the corresponding
chemical connectors near an interface. These will be helpful in Langevin
simulations of autophoretic particles in various experimentally realisable
settings and for studying fluctuating trajectories of an active particle
including both chemical and hydrodynamic interactions.

Aside from its intrinsic theoretical significance, the single-body
solution (exact away from boundaries and approximate in complex environments)
holds potential value in numerically solving the boundary-domain integral
equation for multiple particles. This is due to the ability to initiate
numerical iterations with the single-body solution. For problems falling
under this category, discretised versions of the boundary-domain integral
equations result in diagonally dominant linear systems. Notably, the
one-body solution serves as the solution in cases where hydrodynamic
interactions are disregarded. This implies that starting iterations
from the one-body solution can lead to rapid convergence towards diagonally
dominant numerical solutions \citep{singhGeneralizedStokesLaws2018}.
In scenarios involving multiple interacting particles, utilising a
basis of tensor spherical harmonics for expanding surface fields offers
distinct advantages over other bases like spherical or vector spherical
harmonics, including reduced computational cost due to covariance
under rotations \citep{greengardFastAlgorithmParticle1987,damourMultipoleAnalysisElectromagnetism1991,applequistMaxwellCartesianSpherical2002,turkModellingInferenceActive2023}.
The condition for tangential slip flow in terms of TSH in Eq. (\ref{eq:mode-locking})
now connects in a straightforward way the formalism for general slip
(restricted by mass conservation only) used in previous works \citep{ghoseIrreducibleRepresentationsOscillatory2014,singhManybodyMicrohydrodynamicsColloidal2015,singhGeneralizedStokesLaws2018,singhCompetingChemicalHydrodynamic2019,turkStokesTractionActive2022}
to the present and other problems in which tangential slip is considered,
e.g. active drops. 

In future work we will analytically and numerically build upon the
theoretical results contained in this paper. A detailed analysis of
the dynamical system in Eq. (\ref{eq:EoM}) governing autophoresis
near an interface might reveal features such as intricate, thermally
limited bound states \citep{bolithoPeriodicOrbitsActive2020,mozaffariSelfpropelledColloidalParticle2018} 
potentially relevant to the study of biofilm formation in bacteria
\citep{wilkingBiofilmsComplexFluids2011,persatMechanicalWorldBacteria2015}.
Removing the assumption of rapid diffusion gives rise to nonlinear
advection-diffusion coupling, uncovering a range of potential applications
such as the intricate dynamics of active droplets \citep{herminghaus2014interfacial,michelinSelfpropulsionChemicallyactiveDroplets2023}.
These and more provide exciting avenues for future research.
\begin{acknowledgments}
We thank Professors M. E. Cates, I. Pagonabarraga, and H. A. Stone
for many helpful discussions. We also thank two anonymous referees
for their feedback and constructive criticism, which led to an improvement
in the presentation of our results. G.T. was funded in part by NSF
through the Princeton University (PCCM) Materials Research Science
and Engineering Center DMR-2011750 (co-PI is H. A. Stone), a David
Crighton Fellowship by the Department of Applied Mathematics and Theoretical
Physics at the University of Cambridge to conduct research in the
Department of Physics at the Indian Institute of Technology, Madras,
India, and by the Engineering and Physical Sciences Research Council
(project Reference No. 2089780). R.S. acknowledges support from the Indian Institute of Technology, Madras, India and their seed and initiation grants as well as a Start-up Research Grant, SERB, India (SERB File Number: SRG/2022/000682).

\noindent \textbf{Declaration of Interests.} The authors report no
conflict of interest.
\end{acknowledgments}

\appendix
%dummy comment inserted by tex2lyx to ensure that this paragraph is not empty%%--------------------------------
%%--------------------------------

\section{Chemical problem\label{sec:Chemical-problem}}

\subsection{Exact solution for integral equations\label{subsec:Exact-solution-for}}

As discussed in a previous work \citep{singhCompetingChemicalHydrodynamic2019}
using Galerkin's method, the BIE (IIa) can be expressed as the linear
system 
\begin{equation}
\tfrac{1}{2}\boldsymbol{C}^{(q)}=\boldsymbol{C}^{\infty(q)}+\boldsymbol{\mathcal{H}}^{(q,q')}\odot\boldsymbol{J}^{(q')}+\boldsymbol{\mathcal{L}}^{(q,q')}\odot\boldsymbol{C}^{(q')},\label{eq:chemical-LS}
\end{equation}
with the matrix elements\begin{subequations}\label{eq:mat-elms-laplace}
\begin{equation}
\boldsymbol{\mathcal{H}}^{(q,q')}(\boldsymbol{R},\tilde{\boldsymbol{R}})=\tilde{w}_{q}\tilde{w}_{q'}\int\boldsymbol{Y}^{(q)}(\hat{\boldsymbol{b}})H(\boldsymbol{R}+\boldsymbol{b},\tilde{\boldsymbol{R}}+\boldsymbol{b}')\boldsymbol{Y}^{(q')}(\hat{\boldsymbol{b}}'){\rm d}S{\rm d}S',
\end{equation}
\begin{equation}
\boldsymbol{\mathcal{L}}^{(q,q')}(\boldsymbol{R},\tilde{\boldsymbol{R}})=\tilde{w}_{q}w_{q'}\int\boldsymbol{Y}^{(q)}(\hat{\boldsymbol{b}})\hat{\boldsymbol{b}}'\cdot\boldsymbol{L}(\boldsymbol{R}+\boldsymbol{b},\tilde{\boldsymbol{R}}+\boldsymbol{b}')\boldsymbol{Y}^{(q')}(\hat{\boldsymbol{b}}'){\rm d}S{\rm d}S'.
\end{equation}
\end{subequations}Here, we evaluate these integrals for an unbounded
domain, when $H=H^{o}(\boldsymbol{r})$ (see Eq. (\ref{eq:Green-split}))
and $\boldsymbol{L}=\boldsymbol{L}^{o}(\boldsymbol{r})$, given by
$\boldsymbol{L}^{o}(\boldsymbol{r})=\hat{\boldsymbol{r}}/4\pi r^{2}$.
The matrix elements for the unbounded domain have singular but integrable
kernels. Due to their translational invariance they can be solved
using Fourier techniques. The derivation follows analogous steps to
the one of the exact solution for the Stokes traction for an isolated
active particle in \citep{turkStokesTractionActive2022}. Writing
$\boldsymbol{\mathcal{H}}^{o(q,q')}$ and $\boldsymbol{\mathcal{L}}^{o(q,q')}$
for the corresponding matrix elements, we find
\begin{multline}
\mathcal{H}_{QQ'}^{o(q,q')}=\sum_{n,n'=0}^{\infty}\tau_{nn'qq'}\int{\rm d}SY_{Q}^{(q)}(\hat{\boldsymbol{b}})Y_{N}^{(n)}(\hat{\boldsymbol{b}})\int{\rm d}S'Y_{Q'}^{(q')}(\hat{\boldsymbol{b}}')Y_{N'}^{(n')}(\hat{\boldsymbol{b}}')\\
\times\int{\rm d}kj_{n}(kb)j_{n'}(kb)\int{\rm d}\Omega_{k}Y_{N}^{(n)}(\hat{\boldsymbol{k}})k^{2}\hat{H}^{o}(\boldsymbol{k})Y_{N'}^{(n')}(\hat{\boldsymbol{k}}),
\end{multline}
for the single-layer and similarly for the double-layer,
\begin{multline}
\mathcal{L}_{QQ'}^{o(q,q')}=\sum_{l,l'=0}^{\infty}\tau_{nn'qq'}\int{\rm d}SY_{Q}^{(q)}(\hat{\boldsymbol{b}})Y_{N}^{(n)}(\hat{\boldsymbol{b}})\int{\rm d}S'Y_{Q'}^{(q')}(\hat{\boldsymbol{b}}')Y_{N'}^{(n')}(\hat{\boldsymbol{b}}')\boldsymbol{Y}^{(1)}(\hat{\boldsymbol{b}}')\\
\times\int{\rm d}kkj_{n}(kb)j_{n'}(kb)\int{\rm d}\Omega_{k}Y_{N}^{(n)}(\hat{\boldsymbol{k}})k\hat{\boldsymbol{L}}^{o}(\boldsymbol{k})Y_{N'}^{(n')}(\hat{\boldsymbol{k}}).
\end{multline}
Here, we have defined $\tau_{nn'qq'}=\tfrac{2b^{4}}{\pi}i^{n+3n'}\tilde{w}_{q}\tilde{w}_{q'}w_{n}\tilde{w}_{n}w_{n'}\tilde{w}_{n'}$
and used the Fourier transforms of the Green's functions for the unbounded
domain 
\begin{equation}
\hat{H}^{o}(\boldsymbol{k})=\frac{1}{D}\frac{1}{k^{2}},\qquad\hat{\boldsymbol{L}}^{o}(\boldsymbol{k})=i\frac{\hat{\boldsymbol{k}}}{k}.
\end{equation}
The functions $j_{n}(kb$) are spherical Bessel functions, $b=|\boldsymbol{b}|$
and $i=\sqrt{-1}$ is the imaginary unit. Further, $\int{\rm d}S$
implies an integral over the surface of a sphere with radius $b$,
$\int{\rm d}\Omega$ the integral over the surface of a unit sphere
and $\int{\rm d}k$ a scalar definite integral from $0$ to $\infty$.
Evaluating these expressions, we find that the single- and double-layer
matrix elements diagonalise simultaneously in a basis of TSH such
that
\begin{equation}
\boldsymbol{\mathcal{H}}^{o(q,q')}\odot\boldsymbol{J}^{(q')}=\frac{1}{4\pi bDw_{q}}\boldsymbol{J}^{(q)},\qquad\boldsymbol{\mathcal{L}}^{o(q,q')}\odot\boldsymbol{C}^{(q')}=-\frac{1}{2(2q+1)}\boldsymbol{C}^{(q)}.
\end{equation}
The linear system in Eq. (\ref{eq:chemical-LS}) can then be solved
trivially. We find the exact result, valid for an arbitrary mode index
$q$, 
\begin{equation}
\boldsymbol{C}^{(q)}=\zeta_{q}\boldsymbol{C}^{\infty(q)}+\mathcal{E}_{q}\boldsymbol{J}^{(q)}.\label{eq:exact-conc}
\end{equation}
with $\zeta_{q}$ and $\mathcal{E}_{q}$ given in Eq. (\ref{eq:exact-conc}).
In deriving this result, we corrected an error in the double-layer
calculation given in \citep{singhCompetingChemicalHydrodynamic2019}.

For the matrix elements due to additional boundary conditions with
the propagator $H^{*}$ and the corresponding double-layer $L^{*}$
it is known that Eqs. (\ref{eq:mat-elms-laplace}) evaluate to \citep{singhCompetingChemicalHydrodynamic2019}
\begin{gather}
\boldsymbol{\mathcal{H}}^{*(q,q')}=b^{q+q'}\boldsymbol{\nabla}^{(q)}\tilde{\boldsymbol{\nabla}}^{(q')}H^{*}(\boldsymbol{R},\tilde{\boldsymbol{R}}),\qquad\boldsymbol{\mathcal{L}}^{*(q,q')}=\frac{4\pi bD}{(q'-1)!(2q'+1)!!}\boldsymbol{\mathcal{H}}^{*(q,q')},\label{eq:matelms}
\end{gather}
where we have left the point of evaluation, $\boldsymbol{R}=\tilde{\boldsymbol{R}}$
for the one-body problem, implicit for brevity, and where $\boldsymbol{\mathcal{L}}^{*(q,q')}$
is defined for $q'\geq1$. 

\subsection{Iterative solution in complex environments\label{subsec:Iterative-solutions}}

The formal solution of the boundary-domain integral equation for the
concentration field in Eq. (IId) in a basis of TSH gives the following
for the linear response to a background concentration field,
\begin{equation}
\boldsymbol{\zeta}^{(q,q')}=\left[\boldsymbol{A}^{-1}\right]^{(q,q')},\qquad{\rm where}\qquad\boldsymbol{A}^{(q,q')}=\left(\tfrac{1}{2}\boldsymbol{I}-\boldsymbol{\mathcal{L}}\right)^{(q,q')}.
\end{equation}
This can be computed using Jacobi's iterative method of matrix inversion.
At the $n$-th iteration, we find
\begin{equation}
\left(\boldsymbol{\zeta}^{(q,q')}\right)^{[n]}=\left(\boldsymbol{A}^{(q,q)}\right)^{-1}\Big[\boldsymbol{I}^{(q,q')}-\sum^{'}\boldsymbol{A}^{(q,q'')}\left(\boldsymbol{\zeta}^{(q'',q')}\right)^{[n-1]}\Big],\quad{\rm with}\quad\left(\boldsymbol{\zeta}^{(q,q')}\right)^{[0]}=\zeta_{q}\boldsymbol{I}^{(q,q')}.\label{eq:extC-jacobi}
\end{equation}
The primed sum implies that diagonal terms with $q=q''$ are not included.
Naturally, we choose the solution in the unbounded domain as the zeroth
order solution. Similarly, it is known that at the $n$-th iteration
the elastance in a basis of TSH is given by \citep{singhCompetingChemicalHydrodynamic2019}
\begin{equation}
\left(\boldsymbol{\mathcal{E}}^{(q,q')}\right)^{[n]}=\left(\boldsymbol{A}^{(q,q)}\right)^{-1}\Big[\boldsymbol{\mathcal{H}}^{(q,q')}-\sum^{'}\boldsymbol{A}^{(q,q'')}\left(\boldsymbol{\mathcal{E}}^{(q'',q')}\right)^{[n-1]}\Big],\quad{\rm with}\quad\left(\boldsymbol{\mathcal{E}}^{(q,q')}\right)^{[0]}=\mathcal{E}_{q}\boldsymbol{I}^{(q,q')}.\label{eq:elastance-jacobi}
\end{equation}
To first order in the iteration this yields the expressions given
in Eqs. (\ref{eq:zetaqq'}) and (\ref{eq:epsqq'}), with an error
$\mathcal{O}_{a}$ given in Table \ref{tab:bigO}. 

\subsection{Chemo-hydrodynamic coupling\label{sec:Chemohydrodynamic-coupling}}

The chemo-hydrodynamic coupling tensors in a basis of TSH in Eqs.
(Ih) and (\ref{eq:slip-modes}) are in general given by the surface
integral
\begin{equation}
\boldsymbol{\chi}^{(l,q)}=\tfrac{1}{b}\tilde{w}_{l-1}\sum_{q'=0}^{\infty}w_{q}\tilde{w}_{q'}\boldsymbol{M}^{(q')}\int\left\{ (q+1)\boldsymbol{Y}^{(1)}\boldsymbol{Y}^{(l-1)}\boldsymbol{Y}^{(q')}\boldsymbol{Y}^{(q)}-\boldsymbol{Y}^{(l-1)}\boldsymbol{Y}^{(q')}\boldsymbol{Y}^{(q+1)}\right\} {\rm d}S.\label{eq:chi-coupling}
\end{equation}
For the leading polar, chiral and symmetric modes of slip we have
evaluated them in Eq. (\ref{eq:rel-slip-modes}).  This corrects
a previously erroneous result \citep{singhCompetingChemicalHydrodynamic2019}. 

\section{Hydrodynamic problem and rigid-body motion\label{sec:Hydrodynamic-problem--}}

In the following we include the rigid-body motion of the particle,
$\boldsymbol{v}^{\mathcal{D}}(\boldsymbol{b})=\boldsymbol{V}+\boldsymbol{\Omega}\times\boldsymbol{b}$,
in the expansion in Eq. (\ref{eq:exp-vs}) such that $\boldsymbol{V}^{(1s)}=\boldsymbol{V}-\boldsymbol{V}^{\mathcal{A}}$
and $\boldsymbol{V}^{(2a)}/2b=\boldsymbol{\Omega}-\boldsymbol{\Omega}^{\mathcal{A}}$
for simplicity of notation. As discussed in previous work using a
Galerkin method \citep{singhManybodyMicrohydrodynamicsColloidal2015},
the BIE (IIf) for Stokes flow without thermal fluctuations can be
expressed as the linear system 

\begin{equation}
\tfrac{1}{2}\boldsymbol{V}^{(l\sigma)}=-\boldsymbol{\mathcal{G}}^{(l\sigma,l^{\prime}\sigma^{\prime})}\odot\boldsymbol{F}^{(l^{\prime}\sigma^{\prime})}+\boldsymbol{\mathcal{K}}^{(l\sigma,l^{\prime}\sigma^{\prime})}\odot\boldsymbol{V}^{(l^{\prime}\sigma^{\prime})}.\label{eq:lin-system}
\end{equation}
The matrix elements corresponding to the single- and double layer
integrals are \begin{subequations}\label{eq:matrix-element-def}
\begin{align}
\boldsymbol{\mathcal{G}}^{(l,l^{\prime})}(\boldsymbol{R},\tilde{\boldsymbol{R}}) & =\tilde{w}_{l-1}\tilde{w}_{l'-1}\int\boldsymbol{Y}^{(l-1)}(\hat{\boldsymbol{b}})\boldsymbol{G}(\boldsymbol{R}+\boldsymbol{b},\tilde{\boldsymbol{R}}+\boldsymbol{b}^{\prime})\boldsymbol{Y}^{(l^{\prime}-1)}(\hat{\boldsymbol{b}}^{\prime}){\rm d}S{\rm d}S^{\prime},\label{eq:matrix-element-def-G}\\
\boldsymbol{\mathcal{K}}^{(l,l^{\prime})}(\tilde{\boldsymbol{R}},\boldsymbol{R}) & =\tilde{w}_{l-1}w_{l'-1}\int\boldsymbol{Y}^{(l-1)}(\hat{\boldsymbol{b}})\boldsymbol{K}(\tilde{\boldsymbol{R}}+\boldsymbol{b}^{\prime},\boldsymbol{R}+\boldsymbol{b})\cdot\hat{\boldsymbol{b}}^{\prime}\boldsymbol{Y}^{(l^{\prime}-1)}(\hat{\boldsymbol{b}}^{\prime}){\rm d}S{\rm d}S^{\prime}.\label{eq:matrix-element-def-K}
\end{align}
\end{subequations}In defining the double-layer matrix element it
is worthwhile noting the following. Both double-layer integrals (IIc)
and (IIh) in Table \ref{tab:CH-int} are defined as improper integrals
when $\boldsymbol{r}\in S$, usually referred to as the principal
value. This definition differs from the Cauchy principal value of
a singular one-dimensional integral. While the latter requires excluding
small intervals around the singularity and taking the limit as their
size tends to zero simultaneously, the double-layer integrals both
are weakly singular (given $S$ is a Lyapunov surface), and so their
principal value exists in the usual sense of an improper integral
and is a continuous function in $\boldsymbol{r}\in S$ \citep{pozrikidisBoundaryIntegralSingularity1992,kim2005}. 

Writing the matrix elements as a sum of unbounded and correction terms,
it is known that they evaluate to \citep{singhManybodyMicrohydrodynamicsColloidal2015,turkStokesTractionActive2022}\begin{subequations}\label{eq:matrix-elements}
\begin{align}
\boldsymbol{\mathcal{G}}^{(l,l^{\prime})} & =\boldsymbol{\mathcal{G}}^{o(l,l^{\prime})}+\boldsymbol{\mathcal{G}}^{*(l,l^{\prime})}=\boldsymbol{\mathcal{G}}^{o(l,l^{\prime})}+b^{l+l^{\prime}-2}\mathcal{F}^{l-1}\tilde{\mathcal{F}}^{l^{\prime}-1}\boldsymbol{\nabla}^{(l-1)}\tilde{\boldsymbol{\nabla}}^{(l^{\prime}-1)}\boldsymbol{G}^{*}(\boldsymbol{R},\tilde{\boldsymbol{R}}),\label{eq:MatrixElement-G}\\
\boldsymbol{\mathcal{K}}^{(l,l^{\prime})} & =\boldsymbol{\mathcal{K}}^{o(l,l^{\prime})}+\boldsymbol{\mathcal{K}}^{*(l,l^{\prime})}\nonumber \\
 & =\boldsymbol{\mathcal{K}}^{o(l,l^{\prime})}+\frac{4\pi b^{l+l^{\prime}-1}}{(l^{\prime}-2)!(2l^{\prime}-1)!!}\mathcal{F}^{l-1}\tilde{\mathcal{F}}^{l^{\prime}-1}\boldsymbol{\nabla}^{(l-1)}\tilde{\boldsymbol{\nabla}}^{(l^{\prime}-2)}\boldsymbol{K}^{*}(\tilde{\boldsymbol{R}},\boldsymbol{R}).\label{eq:MatrixElement-K}
\end{align}
\end{subequations}These expressions are exact for a spherical particle.

Defining the column vectors for the force and torque acting on the
particle $\boldsymbol{F}^{A}=\left(\boldsymbol{F}^{(1s)},\boldsymbol{F}^{(2a)}\right)^{\text{tr}},$the
higher moments of traction $\boldsymbol{F}^{B}=\left(\boldsymbol{F}^{(2s)},\boldsymbol{F}^{(3t)},\dots\right)^{\text{tr}}$,
the modes corresponding to rigid-body motion $\boldsymbol{V}^{A}=\left(\boldsymbol{V}^{(1s)},\boldsymbol{V}^{(2a)}\right)^{\text{tr}} $
and the higher modes of the slip $\boldsymbol{V}^{B}=\left(\boldsymbol{V}^{(2s)},\boldsymbol{V}^{(3t)},\dots\right)^{\text{tr}}$,
we can write the linear system as \citep{singhManybodyMicrohydrodynamicsColloidal2015}

\begin{equation}
\frac{1}{2}\begin{pmatrix}\boldsymbol{V}^{A}\\
\boldsymbol{V}^{B}
\end{pmatrix}=-\begin{pmatrix}\boldsymbol{\mathcal{G}}^{AA} & \boldsymbol{\mathcal{G}}^{AB}\\
\boldsymbol{\mathcal{G}}^{BA} & \boldsymbol{\mathcal{G}}^{BB}
\end{pmatrix}\begin{pmatrix}\boldsymbol{F}^{A}\\
\boldsymbol{F}^{B}
\end{pmatrix}+\begin{pmatrix}\boldsymbol{\mathcal{K}}^{AA} & \boldsymbol{\mathcal{K}}^{AB}\\
\boldsymbol{\mathcal{K}}^{BA} & \boldsymbol{\mathcal{K}}^{BB}
\end{pmatrix}\begin{pmatrix}\boldsymbol{V}^{A}\\
\boldsymbol{V}^{B}
\end{pmatrix}.\label{eq:block-lin-system}
\end{equation}
To be able to solve this infinite linear system, we need to truncate
the mode expansions (\ref{eq:exp-vs}) at some appropriate order,
and fix the gauge freedom in the traction. Taking care of the latter,
we impose $\int\boldsymbol{f}^{H}\cdot\hat{\boldsymbol{b}}{\rm d}S=\,-\int p{\rm d}S=0$,
which is equivalent to imposing $F^{(2t)}=0$. The rationale behind
this can be explained as follows. The pressure is a harmonic function,
i.e., $\nabla^{2}p=0$, and can thus be expanded in a basis constructed
from derivatives of $1/r$. The leading mode of such an expansion
decays as $1/r$ and its expansion coefficient is obtained from the
integral $\int p\,{\rm d}S$. Further, incompressibility, and the
absence of sinks and sources of fluid render the pressure a non-dynamical
quantity, meaning that the fundamental solution for the fluid flow
$\boldsymbol{v}$ is independent of the pressure and decays as $1/r$.
However, Stokes equation (Id) must still be satisfied, and a pressure
term decaying as $1/r$ would violate it. We thus impose $\int p\,\mathrm{d}S=0$,
rendering the single-layer operator invertible. Eliminating the unknown
$\boldsymbol{F}^{B}$, we can directly solve for the rigid-body motion
of the particle 
\begin{equation}
\boldsymbol{V}^{A}=-\mathbb{M}\cdot\boldsymbol{F}^{A}+\boldsymbol{\Pi}\odot\boldsymbol{V}^{B},\label{eq:formal-solution}
\end{equation}
where we have defined the grand mobility matrix $\mathbb{M}$ and
the grand propulsion tensor $\boldsymbol{\Pi}$,
\begin{gather}
\mathbb{M}=\left[\boldsymbol{\mathcal{G}}^{AA}-\boldsymbol{\mathcal{G}}^{AB}\left(\boldsymbol{\mathcal{G}}^{BB}\right)^{-1}\boldsymbol{\mathcal{G}}^{BA}\right],\qquad\boldsymbol{\Pi}=\left[\boldsymbol{\mathcal{K}}^{AB}+\boldsymbol{\mathcal{G}}^{AB}\left(\boldsymbol{\mathcal{G}}^{BB}\right)^{-1}\left(\tfrac{1}{2}\boldsymbol{I}-\boldsymbol{\mathcal{K}}^{BB}\right)\right].\label{eq:grand-tensors}
\end{gather}
In finding this solution, we have used that rigid-body motion lies
in the eigenspace of the double layer matrix element with a uniform
eigenvalue of $-1/2$, and that no exterior flows are produced for
the rigid-body component of the motion such that 
\begin{equation}
\boldsymbol{\mathcal{K}}^{AA}\boldsymbol{V}^{A}=-\tfrac{1}{2}\boldsymbol{V}^{A},\quad\quad\boldsymbol{\mathcal{K}}^{BA}\mathbf{V}^{A}=\mathbf{0}.
\end{equation}
Equation \ref{eq:grand-tensors} guarantees a positive-definite mobility
matrix given that every principal sub-matrix of a positive-definite
matrix (here, $\boldsymbol{\mathcal{G}}^{(l\sigma,l^{\prime}\sigma^{\prime})}$)
is positive-definite itself. Comparing Eqs. (\ref{eq:EoM}) and (\ref{eq:grand-tensors})
we can directly identify the mobility and propulsion tensors in terms
of the matrix elements in Eq. (\ref{eq:matrix-element-def}). For
the mobilities we find

\begin{equation}
\boldsymbol{\mu}^{\alpha\beta}=\frac{1}{c^{\alpha}q^{\beta}}\Big[\boldsymbol{\mathcal{G}}^{(l\sigma,l^{\prime}\sigma^{\prime})}-\sum_{l^{\prime\prime}\sigma^{\prime\prime}=2s}\boldsymbol{\mathcal{G}}^{(l\sigma,l^{\prime\prime}\sigma^{\prime\prime})}\boldsymbol{\Upsilon}{}^{(l^{\prime\prime}\sigma^{\prime\prime},l^{\prime}\sigma^{\prime})}\Big],\label{eq:mobility}
\end{equation}
with $\alpha=T,R$ implying $l\sigma=1s,2a$ and $\beta=T,R$ implying
$l^{\prime}\sigma^{\prime}=1s,2a$, respectively. The scalar pre-factors
$c^{\alpha}$ and $q^{\beta}$ can be found in Table \ref{tab:cq}.
\begin{table}
\begin{tabular}{>{\centering}p{1.5cm}>{\centering}p{1.5cm}>{\centering}p{1.5cm}}
\toprule 
$\alpha,\beta$ & $T$ & $R$\tabularnewline
\midrule
\midrule 
$c^{\alpha}$ & $1$ & $2b$\tabularnewline
\midrule 
$q^{\beta}$ & $1$ & $b$\tabularnewline
\bottomrule
\end{tabular}\caption{\label{tab:cq}Coefficients $c^{\alpha}$ and $q^{\beta}$ in the
mobility and propulsion tensors.}
\end{table}
 Similarly, we find for the propulsion tensors
\begin{equation}
\boldsymbol{\pi}^{(\alpha,l^{\prime}\sigma^{\prime})}=\frac{1}{c^{\alpha}}\Big[\boldsymbol{\mathcal{K}}^{(l\sigma,l^{\prime}\sigma^{\prime})}+\sum_{l^{\prime\prime}\sigma^{\prime\prime}=2s}\boldsymbol{\mathcal{G}}^{(l\sigma,l^{\prime\prime}\sigma^{\prime\prime})}\boldsymbol{\Phi}^{(l^{\prime\prime}\sigma^{\prime\prime},l^{\prime}\sigma^{\prime})}\Big].\label{eq:propulsion}
\end{equation}
The propulsion tensors are defined for $l^{\prime}\sigma^{\prime}\geq2s$
as follows directly from the equations of motion \eqref{eq:EoM}.
In Eqs. \eqref{eq:mobility} and \eqref{eq:propulsion} we have defined\begin{subequations}
\begin{align}
\boldsymbol{\Upsilon}^{(l\sigma,l^{\prime}\sigma^{\prime})} & =\sum_{l^{\prime\prime}\sigma^{\prime\prime}=2s}\left(\boldsymbol{\mathcal{G}}^{(l\sigma,l^{\prime\prime}\sigma^{\prime\prime})}\right)^{-1}\boldsymbol{\mathcal{G}}^{(l^{\prime\prime}\sigma^{\prime\prime},l^{\prime}\sigma^{\prime})},\\
\boldsymbol{\Phi}^{(l\sigma,l^{\prime}\sigma^{\prime})} & =\sum_{l^{\prime\prime}\sigma^{\prime\prime}=2s}\left(\boldsymbol{\mathcal{G}}^{(l\sigma,l^{\prime\prime}\sigma^{\prime\prime})}\right)^{-1}\left(\tfrac{1}{2}\boldsymbol{I}-\boldsymbol{\mathcal{K}}\right)^{(l^{\prime\prime}\sigma^{\prime\prime},l^{\prime}\sigma^{\prime})}=\sum_{l^{\prime\prime}\sigma^{\prime\prime}=2s}\left(\boldsymbol{\mathcal{G}}^{(l\sigma,l^{\prime\prime}\sigma^{\prime\prime})}\right)^{-1}\mathcal{\boldsymbol{B}}^{(l^{\prime\prime}\sigma^{\prime\prime},l^{\prime}\sigma^{\prime})}.\label{eq:higher-friction}
\end{align}
\end{subequations}Using Jacobi's method of matrix inversion, we find
iterative solutions for the mobility and propulsion tensors. At the
$n$-th iteration we obtain\begin{subequations}\label{eq:mob-prop}
\begin{align}
\left(\boldsymbol{\mu}^{\alpha\beta}\right)^{[n]} & =\frac{1}{c^{\alpha}q^{\beta}}\Big[\boldsymbol{\mathcal{G}}^{(l\sigma,l^{\prime}\sigma^{\prime})}-\sum_{l^{\prime\prime}\sigma^{\prime\prime}=2s}\boldsymbol{\mathcal{G}}^{(l\sigma,l^{\prime\prime}\sigma^{\prime\prime})}\left(\boldsymbol{\Upsilon}^{(l^{\prime\prime}\sigma^{\prime\prime},l^{\prime}\sigma^{\prime})}\right)^{[n]}\Big],\label{eq:mobility-iter}\\
\left(\boldsymbol{\pi}^{(\alpha,l^{\prime}\sigma^{\prime})}\right)^{[n]} & =\frac{1}{c^{\alpha}}\Big[\boldsymbol{\mathcal{K}}^{(l\sigma,l^{\prime}\sigma^{\prime})}+\sum_{l^{\prime\prime}\sigma^{\prime\prime}=2s}\boldsymbol{\mathcal{G}}^{(l\sigma,l^{\prime\prime}\sigma^{\prime\prime})}\left(\boldsymbol{\Phi}^{(l^{\prime\prime}\sigma^{\prime\prime},l^{\prime}\sigma^{\prime})}\right)^{[n]}\Big],\label{eq:propulsion-iter}
\end{align}
\end{subequations}with\begin{subequations}\label{eq:HJ}
\begin{align}
\left(\boldsymbol{\Upsilon}^{(l\sigma,l^{\prime}\sigma^{\prime})}\right)^{[n]} & =\left(\boldsymbol{\mathcal{G}}^{(l\sigma,l\sigma)}\right)^{-1}\Big[\boldsymbol{\mathcal{G}}^{(l\sigma,l^{\prime}\sigma^{\prime})}-\sum_{l^{\prime\prime}\sigma^{\prime\prime}=2s}^{\prime}\boldsymbol{\mathcal{G}}^{(l\sigma,l^{\prime\prime}\sigma^{\prime\prime})}\left(\boldsymbol{\Upsilon}^{(l^{\prime\prime}\sigma^{\prime\prime},l^{\prime}\sigma^{\prime})}\right)^{[n-1]}\Big],\\
\left(\Phi^{(l\sigma,l^{\prime}\sigma^{\prime})}\right)^{[n]} & =\left(\boldsymbol{\mathcal{G}}^{(l\sigma,l\sigma)}\right)^{-1}\Big[\boldsymbol{\mathcal{B}}^{(l\sigma,l^{\prime}\sigma^{\prime})}-\sum_{l^{\prime\prime}\sigma^{\prime\prime}=2s}^{\prime}\boldsymbol{\mathcal{G}}^{(l\sigma,l^{\prime\prime}\sigma^{\prime\prime})}\left(\boldsymbol{\Phi}^{(l^{\prime\prime}\sigma^{\prime\prime},l^{\prime}\sigma^{\prime})}\right)^{[n-1]}\Big].
\end{align}
\end{subequations}The primed sum implies that the diagonal terms
with $l\sigma=l''\sigma''$ are not included. Without loss of generality,
we choose the zeroth order solutions to be
\begin{gather}
\left(\boldsymbol{\Upsilon}^{(l\sigma,l^{\prime}\sigma^{\prime})}\right)^{[0]}=0,\qquad\left(\boldsymbol{\Phi}^{(l\sigma,l^{\prime}\sigma^{\prime})}\right)^{[0]}=\hat{\gamma}_{l\sigma}\,\boldsymbol{I}^{(l\sigma,l'\sigma')},\label{eq:mobility-iter-1}
\end{gather}
where the scalar friction coefficients $\hat{\gamma}_{l\sigma}$ are
given in Ref. \citep{turkStokesTractionActive2022} and $\boldsymbol{I}^{(l\sigma,l'\sigma')}$
is the identity tensor with elements $\delta_{ll'}\delta_{\sigma\sigma'}$.
It is worthwhile to note that with this choice, the iteration at zeroth
order for the mobility and propulsion tensors corresponds to a superposition
approximation, ignoring higher order hydrodynamic interactions. This
yields the expressions in Eqs. (\ref{eq:mobilityleading}) and (\ref{eq:propulsionleading}).
\begin{table}
\centering{}%
\begin{tabular}{>{\centering}p{5.2cm}>{\centering}p{5.2cm}>{\centering}p{5.2cm}}
\toprule 
\addlinespace
$\mathcal{O}_{a}$ & $\mathcal{O}_{b}$ & $\mathcal{O}_{c}$\tabularnewline\addlinespace
\midrule 
\addlinespace
$\mathcal{O}\left(\boldsymbol{\mathcal{L}}^{*(q,q'')}\odot\boldsymbol{\mathcal{H}}^{*(q'',q')}\right)$ & $\mathcal{O}\left(\nabla^{2}\left(\tilde{\boldsymbol{\nabla}}\boldsymbol{G}^{*}\right)\colon\left(\boldsymbol{\nabla}\boldsymbol{G}^{*}\right)\right)$ & $\mathcal{O}\left(\nabla^{2}\left(\tilde{\boldsymbol{\nabla}}\boldsymbol{G}^{*}\right)\colon\left(\boldsymbol{\nabla}\left(\tilde{\boldsymbol{\nabla}}\times\boldsymbol{G}^{*}\right)\right)\right)$\tabularnewline\addlinespace
\midrule
\midrule 
\addlinespace
$\mathcal{O}_{d}$ & $\mathcal{O}_{e}$ & $\mathcal{O}_{f}$\tabularnewline\addlinespace
\midrule 
\addlinespace
$\mathcal{O}\left(\nabla^{2}\left(\tilde{\boldsymbol{\nabla}}\boldsymbol{G}^{*}\right)\colon\left(\boldsymbol{\nabla}\tilde{\boldsymbol{\nabla}}\boldsymbol{G}^{*}\right)\right)$ & $\mathcal{O}\left(\left(\tilde{\boldsymbol{\nabla}}\boldsymbol{G}^{*}\right)\colon\left(\boldsymbol{\nabla}\tilde{\boldsymbol{\nabla}}^{(2)}\cdot\boldsymbol{G}^{*}\right)\right)$ & $\mathcal{O}\left(\left(\tilde{\boldsymbol{\nabla}}\boldsymbol{G}^{*}\right)\colon\left(\boldsymbol{\nabla}\tilde{\boldsymbol{\nabla}}^{(3)}\cdot\boldsymbol{G}^{*}\right)\right)$\tabularnewline\addlinespace
\bottomrule
\end{tabular}\caption{\label{tab:bigO}Big O notation for errors in the linear response
to elastance, mobility and propulsion tensors.}
\end{table}
 Evaluated for a plane interface, they correspond to the mobility
and propulsion coefficients given in Table \ref{tab:Mobility-coefficients}. 

For the exact mobilities and propulsion tensors we can write
\begin{equation}
\boldsymbol{\mu}^{\alpha\beta}=\left(\boldsymbol{\mu}^{\alpha\beta}\right)^{[0]}+\Delta\boldsymbol{\mu}^{\alpha\beta},\qquad\boldsymbol{\pi}^{(\alpha,l\sigma)}=\left(\boldsymbol{\pi}^{(\alpha,l\sigma)}\right)^{[0]}+\Delta\boldsymbol{\pi}^{(\alpha,l\sigma)},
\end{equation}
where the zeroth order terms are given in the main text and, explicit
to leading order, the corrections are
\begin{gather}
\Delta\boldsymbol{\mu}^{TT}=-\tfrac{10\pi\eta b^{3}}{3}\left[\tilde{\boldsymbol{\nabla}}\boldsymbol{G}^{*}+(\tilde{\boldsymbol{\nabla}}\boldsymbol{G}^{*})^{\text{tr}}\right]\colon\boldsymbol{\nabla}\boldsymbol{G}^{*}+\mathcal{O}_{b},\qquad\Delta\boldsymbol{\mu}^{TR}=-\tfrac{5\pi\eta b^{3}}{3}\left[\tilde{\boldsymbol{\nabla}}\boldsymbol{G}^{*}+(\tilde{\boldsymbol{\nabla}}\boldsymbol{G}^{*})^{\text{tr}}\right]\colon\boldsymbol{\nabla}(\tilde{\boldsymbol{\nabla}}\times\boldsymbol{G}^{*})+\mathcal{O}_{c},
\end{gather}
for the mobilities and
\begin{gather}
\Delta\boldsymbol{\pi}^{(T,2s)}=-b\left(\tfrac{10\pi\eta b^{2}}{3}\right)^{2}\left[\tilde{\boldsymbol{\nabla}}\boldsymbol{G}^{*}+(\tilde{\boldsymbol{\nabla}}\boldsymbol{G}^{*})^{\text{tr}}\right]\colon\boldsymbol{\nabla}\left[\tilde{\boldsymbol{\nabla}}\boldsymbol{G}^{*}+(\tilde{\boldsymbol{\nabla}}\boldsymbol{G}^{*})^{\text{tr}}\right]+\mathcal{O}_{d},\nonumber \\
\Delta\boldsymbol{\pi}^{(T,3t)}=\mathcal{O}_{e},\qquad\Delta\boldsymbol{\pi}^{(T,4t)}=\mathcal{O}_{f},
\end{gather}
for the propulsion tensors. Here, a colon indicates a contraction
of two pairs of indices. The higher-order corrections denoted by $\mathcal{O}$
are specified in Table \ref{tab:bigO}. Using Eqs. \ref{eq:mob-prop}
these higher-order terms can be computed to arbitrary accuracy. However,
this is a non-trivial computation and will be the topic of future
work. Evaluated for a plane interface, the leading-order correction
to the mobilities contain the order-$\hat{h}^{-4}$ terms
\begin{equation}
\Delta\mu_{\parallel}^{TT}\big(\hat{h}^{-4}\big)=-\frac{45\lambda^{f}{}^{2}}{256\left(1+\lambda^{f}\right)^{2}},\qquad\Delta\mu_{\perp}^{TT}\big(\hat{h}^{-4}\big)=-\frac{15\left(2+3\lambda^{f}\right)^{2}}{256\left(1+\lambda^{f}\right)^{2}},
\end{equation}
matching previous results in the literature for the special cases
of a wall and a free surface \citep{goldmanSlowViscousMotion,perkinsHydrodynamicInteractionSpherical1990,perkinsHydrodynamicInteractionSpherical1992}.
While it might be tempting to include these next-to-leading-order
coefficients in the results for the mobilities in Table \ref{tab:Mobility-coefficients},
one sacrifices positive-definiteness of the mobility matrix $\mathbb{M}$
if doing so and Brownian simulations can no longer be guaranteed to
work correctly. Positive-definiteness beyond the zeroth iteration
can only be guaranteed at the full first-order Jacobi iteration. In
the case of the propulsion tensors, at order $\hat{h}^{-5}$ the following
terms arise:
\begin{gather}
\Delta\pi_{1}^{(T,2s)}\big(\hat{h}^{-5}\big)=\frac{25\,\lambda^{f}\left(1+3\lambda^{f}\right)}{256\left(1+\lambda^{f}\right)^{2}},\qquad\Delta\pi_{2}^{(T,2s)}\big(\hat{h}^{-5}\big)=-\frac{25\left(2+7\lambda^{f}+6\lambda^{f}{}^{2}\right)}{384\left(1+\lambda^{f}\right)^{2}}.
\end{gather}

\section{Coupling to an interface\label{sec:Simulation}}

Here, we give a detailed account of the simulation of Eqs. (\ref{eq:EoM})
presented in Section \ref{subsec:interface-effect} for a bottom-heavy
Brownian Janus particle near a plane interface. Using the mobilities
for a spherical particle near a plane boundary in Table \ref{tab:Mobility-coefficients},
we find the only non-vanishing convective term to be proportional
to
\begin{equation}
\partial_{z}\mu_{\perp}^{TT}=\frac{1}{6\pi\eta b^{2}}\left[\frac{3(2+3\lambda^{f})}{16(1+\lambda^{f})}\hat{h}^{-2}-\frac{3\left(1+4\lambda^{f}\right)}{16(1+\lambda^{f})}\hat{h}^{-4}+\frac{5\lambda^{f}}{16(1+\lambda^{f})}\hat{h}^{-6}\right],\label{eq:dzmuTT}
\end{equation}
contributing to the dynamics of the particle in $z$-direction which
is to be included in the spurious drift. 

Next, we give an expression for the noise strength $\propto\sqrt{2k_{B}T\,\mathbb{M}}$
in the update equations (\ref{eq:EoM}) for a Brownian particle close
to a plane interface, for which the diffusion matrix takes a particularly
simple form. Using the definitions for the scalar mobility coefficients
from Table \ref{tab:Mobility-coefficients} we define the following
coefficients:\allowdisplaybreaks\begingroup\setlength{\jot}{1ex}
\begin{gather}
\sqrt{\mu_{\parallel}^{2}}\equiv\sqrt{\left(\mu_{\parallel}^{RR}-\mu_{\parallel}^{TT}\right)^{2}+4\left(\mu^{TR}\right)^{2}},\qquad\sqrt{\mu_{\parallel}^{+}}\equiv\sqrt{\mu_{\parallel}^{RR}+\mu_{\parallel}^{TT}+\sqrt{\mu_{\parallel}^{2}}},\qquad\sqrt{\mu_{\parallel}^{-}}\equiv\sqrt{\mu_{\parallel}^{RR}+\mu_{\parallel}^{TT}-\sqrt{\mu_{\parallel}^{2}}}.
\end{gather}
\endgroup Using these we define the further coefficients\allowdisplaybreaks\begingroup\setlength{\jot}{1ex}
\begin{align}
\sqrt{\mu_{xx}} & \equiv\tfrac{1}{\sqrt{8}\sqrt{\mu_{\parallel}^{2}}}\left[\sqrt{\mu_{\parallel}^{-}}\left(\mu_{\parallel}^{RR}-\mu_{\parallel}^{TT}+\sqrt{\mu_{\parallel}^{2}}\right)+\sqrt{\mu_{\parallel}^{+}}\left(\mu_{\parallel}^{TT}-\mu_{\parallel}^{RR}+\sqrt{\mu_{\parallel}^{2}}\right)\right],\\
\sqrt{\mu_{xe}} & \equiv\tfrac{1}{\sqrt{2}\sqrt{\mu_{\parallel}^{2}}}\,\mu^{TR}\left(\sqrt{\mu_{\parallel}^{+}}-\sqrt{\mu_{\parallel}^{-}}\right),\\
\sqrt{\mu_{e_{x}e_{x}}} & \equiv\tfrac{1}{\sqrt{8}\sqrt{\mu_{\parallel}^{2}}}\left[\mu_{\parallel}^{TT}\left(\sqrt{\mu_{\parallel}^{-}}-\sqrt{\mu_{\parallel}^{+}}\right)+\mu_{\parallel}^{RR}\left(\sqrt{\mu_{\parallel}^{+}}-\sqrt{\mu_{\parallel}^{-}}\right)+\sqrt{\mu_{\parallel}^{2}}\left(\sqrt{\mu_{\parallel}^{+}}+\sqrt{\mu_{\parallel}^{-}}\right)\right].
\end{align}
\endgroup Finally, we have
\begin{equation}
\sqrt{\mathbb{M}}=\begin{pmatrix}\sqrt{\mu_{xx}} & 0 & 0 & 0 & \sqrt{\mu_{xe}} & 0\\
0 & \sqrt{\mu_{xx}} & 0 & -\sqrt{\mu_{xe}} & 0 & 0\\
0 & 0 & \sqrt{\mu_{\perp}^{TT}} & 0 & 0 & 0\\
0 & 0 & -\sqrt{\mu_{xe}} & \sqrt{\mu_{e_{x}e_{x}}} & 0 & 0\\
\sqrt{\mu_{xe}} & 0 & 0 & 0 & \sqrt{\mu_{e_{x}e_{x}}} & 0\\
0 & 0 & 0 & 0 & 0 & \sqrt{\mu_{\perp}^{RR}}
\end{pmatrix},\label{eq:sqrtMu}
\end{equation}
which is straightforward to compute. 

\subsection{Parameterisation}

The update equations can be simplified further by uniaxially parametrising
the slip modes in the propulsion terms in Eq. (\ref{eq:prop-terms}).
We write
\begin{equation}
\boldsymbol{V}^{\mathcal{A}}=V^{\mathcal{A}}\boldsymbol{e}_{{\scriptscriptstyle V}},\qquad\boldsymbol{\Omega}^{\mathcal{A}}=\Omega^{\mathcal{A}}\boldsymbol{e}_{{\scriptscriptstyle \Omega}},\qquad\boldsymbol{V}_{s}^{(2s)}=V_{s}^{(2s)}\left(3\boldsymbol{e}_{{\scriptscriptstyle S}}\boldsymbol{e}_{{\scriptscriptstyle S}}-\boldsymbol{I}\right),\label{eq:uniaxial-slipModes}
\end{equation}
where the strengths $V^{\mathcal{A}}$, $\Omega^{\mathcal{A}}$, and
$V_{s}^{(2s)}$ of the modes and their respective orientations $\boldsymbol{e}_{{\scriptscriptstyle V}}$,
$\boldsymbol{e}_{{\scriptscriptstyle \Omega}}$ and $\boldsymbol{e}_{{\scriptscriptstyle S}}$
are obtained from Eq. (\ref{eq:rel-slip-modes}) and given below.
The leading symmetric mode is defined as $\boldsymbol{V}_{s}^{(2s)}=\frac{3}{8\pi b^{2}}\int\left\{ \hat{\boldsymbol{b}}\boldsymbol{v}^{\mathcal{A}}+(\hat{\boldsymbol{b}}\boldsymbol{v}^{\mathcal{A}})^{\text{tr}}\right\} {\rm d}S.$
For the polar and symmetric modes we define the polar angle $\vartheta_{\alpha}$,
where $\alpha=V,S$, such that

\begin{equation}
\boldsymbol{e}_{\alpha}=\cos\vartheta_{\alpha}\,\hat{\boldsymbol{x}}+\sin\vartheta_{\alpha}\,\hat{\boldsymbol{z}},
\end{equation}
while for motion in the $x$-$z$ plane it follows that $\boldsymbol{e}_{{\scriptscriptstyle \Omega}}=\hat{\boldsymbol{y}}$.
Far away from the interface ($h/b\gg1$) we have $\vartheta_{{\scriptscriptstyle V}}=\vartheta_{{\scriptscriptstyle S}}=\vartheta$.
We assume that the particle is in the positive half-space above the
interface such that $z=h$. This yields the mean translational dynamics
in the $x$-$z$ plane (with no mean translation in the $y$-direction)\begingroup\renewcommand*{\arraystretch}{1.7}
\begin{equation}
\begin{pmatrix}\left\langle \dot{x}\right\rangle \\
\langle\dot{h}\rangle
\end{pmatrix}=\begin{pmatrix}\mu^{TR}\kappa\cos\vartheta\\
-\mu_{\perp}^{TT}\,mg+k_{B}T\partial_{z}\mu_{\perp}^{TT}
\end{pmatrix}+V^{\mathcal{A}}\begin{pmatrix}(1+5\pi_{\parallel}^{(T,3t)})\cos\vartheta_{{\scriptscriptstyle V}}\\
(1+5\pi_{\perp}^{(T,3t)})\sin\vartheta_{{\scriptscriptstyle V}}
\end{pmatrix}+3V_{s}^{(2s)}\begin{pmatrix}(\pi_{1}^{(T,2s)}-14\pi_{1}^{(T,4t)})\sin\left(2\vartheta_{{\scriptscriptstyle S}}\right)\\
(\pi_{2}^{(T,2s)}-14\pi_{2}^{(T,4t)})(1-3\sin^{2}\vartheta_{{\scriptscriptstyle S}})
\end{pmatrix},\label{eq:simT}
\end{equation}
where the thermal contribution arises from the convective term in
the positional update equation (\ref{eq:update-position}) and is
given in Eq. (\ref{eq:dzmuTT}). \endgroup The mean orientational
dynamics are governed by the angular velocity (with $\dot{\vartheta}=-\Omega_{y}$)
\begin{equation}
\left\langle \Omega_{y}\right\rangle =\mu_{\parallel}^{RR}\kappa\cos\vartheta+\Omega^{\mathcal{A}}+5V^{\mathcal{A}}\pi^{(R,3t)}\,\cos\vartheta_{{\scriptscriptstyle V}}+3V_{s}^{(2s)}\left(\pi^{(R,2s)}-14\pi^{(R,4t)}\right)\sin\left(2\vartheta_{{\scriptscriptstyle S}}\right).\label{eq:simR}
\end{equation}
It is important to note that the brackets $\left\langle \cdot\right\rangle $
simply imply that we are not explicitly writing down the noise terms
proportional to Eq. (\ref{eq:sqrtMu}). To find the true average trajectory
at finite temperature one has to extract it from the full positional
and orientational probability distribution functions of the particle.
This is beyond the scope of this paper.

We now define the coefficients in the dynamical system governing autophoresis
in Eqs. (\ref{eq:simT}) and (\ref{eq:simR}) in terms of the phoretic
model parameters of Eq. (\ref{eq:trunc-chem}). We write the vectorial
part of the phoretic mobility and the generated concentration
field components as
\begin{equation}
M_{x}^{(1)}=M_{1}\cos\left(\vartheta+\psi\right),\quad\quad M_{z}^{(1)}=M_{1}\sin\left(\vartheta+\psi\right),\qquad C_{x}^{(1)}=\mathcal{E}_{\parallel}^{(1,1)}J_{1}\cos\vartheta,\quad C_{z}^{(1)}=\mathcal{E}^{(1,0)}J_{0}+\mathcal{E}_{\perp}^{(1,1)}J_{1}\sin\vartheta.\label{eq:M1comps}
\end{equation}
Comparing the parameterisations in Eq. (\ref{eq:uniaxial-slipModes})
to the definition of $\boldsymbol{V}^{\mathcal{A}}$ in Eq. (\ref{eq:rel-slip-modes}),
and $\boldsymbol{\Omega}^{\mathcal{A}}=\Omega^{\mathcal{A}}\hat{\boldsymbol{y}}$
for the angular speed, we obtain for the corresponding terms in the
dynamical system
\begin{align}
V^{\mathcal{A}}\cos\vartheta_{{\scriptscriptstyle V}} & =-\tfrac{1}{6\pi b^{3}}M_{0}C_{x}^{(1)}-\tfrac{3}{20\pi b^{3}}\left(M_{x}^{(1)}\left(\mathcal{E}^{(2,0)}J_{0}-\mathcal{E}^{(2,1)}J_{1}\sin\vartheta\right)+\mathcal{E}^{(2,1)}M_{z}^{(1)}J_{1}\cos\vartheta\right),\\
V^{\mathcal{A}}\sin\vartheta_{{\scriptscriptstyle V}} & =-\tfrac{1}{6\pi b^{3}}M_{0}C_{z}^{(1)}-\tfrac{3}{20\pi b^{3}}\left(\mathcal{E}^{(2,1)}M_{x}^{(1)}J_{1}\cos\vartheta+2M_{z}^{(1)}\left(\mathcal{E}^{(2,1)}J_{1}\sin\vartheta-\mathcal{E}^{(2,0)}J_{0}\right)\right),\\
\Omega^{\mathcal{A}} & =-\tfrac{3}{8\pi b^{4}}\left(M_{z}^{(1)}C_{x}^{(1)}-M_{x}^{(1)}C_{z}^{(1)}\right).
\end{align}
Finally, using the definition of $\boldsymbol{V}_{s}^{(2s)}$ in Eq.
(\ref{eq:rel-slip-modes}), we find
\begin{align}
V_{s}^{(2s)}\sin2\vartheta_{{\scriptscriptstyle S}} & =\tfrac{1}{20\pi b^{3}}\left(3\left(M_{x}^{(1)}C_{z}^{(1)}+M_{z}^{(1)}C_{x}^{(1)}\right)+2\mathcal{E}^{(2,1)}M_{0}J_{1}\cos\vartheta\right),\\
V_{s}^{(2s)}\left(1-3\sin^{2}\vartheta_{{\scriptscriptstyle S}}\right) & =\tfrac{3}{20\pi b^{3}}\left[M_{x}^{(1)}C_{x}^{(1)}-2M_{z}^{(1)}C_{z}^{(1)}+2M_{0}\left(\mathcal{E}^{(2,0)}J_{0}-\mathcal{E}^{(2,1)}J_{1}\sin\vartheta\right)\right].
\end{align}

\subsection{Approximations}

\subsubsection{Unbounded domain:}

In the unbounded domain the mean dynamics simplify to\begin{subequations}\label{eq:unbounded-eom}
\begin{gather}
\left\langle \dot{x}\right\rangle =-\tfrac{1}{6\pi b^{3}}\mathcal{E}_{1}M_{0}J_{1}\cos\vartheta,\quad\langle\dot{h}\rangle=-\mu_{T}\,mg-\tfrac{1}{6\pi b^{3}}\mathcal{E}_{1}M_{0}J_{1}\sin\vartheta,\quad\langle\dot{\vartheta}\rangle=-\mu_{R}\kappa\cos\vartheta+\tfrac{3}{8\pi b^{4}}\mathcal{E}_{1}M_{1}J_{1}\sin\psi,\label{eq:z-unbounded}
\end{gather}
\end{subequations}with $\mathcal{E}_{1}=3/8\pi bD_{1}$. It is clear
that even for a force- and torque-free particle ($g=\kappa=0$) in
an unbounded fluid, autophoretic motion takes place for the model
considered in Eq. (\ref{eq:trunc-chem}). Neglecting bottom-heaviness
of the particle, the average self-propulsion and self-rotation speeds
in an unbounded fluid are 
\begin{equation}
V^{\mathcal{A}}=\tfrac{1}{6\pi b^{3}}\mathcal{E}_{1}M_{0}J_{1},\qquad\Omega^{\mathcal{A}}=\tfrac{3}{8\pi b^{4}}\mathcal{E}_{1}M_{1}J_{1}\sin\left(\psi\right).
\end{equation}

\subsubsection{Far from the interface -- leading-order effects:}

Considering terms up to order $\hat{h}^{-1}$, hydrodynamic interactions
with the boundary are the first to manifest themselves by altering
the unbounded equations (\ref{eq:unbounded-eom}) as follows,

\begin{gather}
\mu_{T}\rightarrow\mu_{T}\left(1-\Lambda_{T}^{f}\hat{h}^{-1}\right),\qquad{\rm with}\qquad\Lambda_{T}^{f}=\frac{3(2+3\lambda^{f})}{8(1+\lambda^{f})}.
\end{gather}
At this order, chemical interactions with the interface do not yet
appear. Compared to the unbounded equations, the orientational and
parallel dynamics are unaffected.

\subsubsection{Far from the interface -- next-to-leading-order effects:}

Considering terms up to $\hat{h}^{-2}$ leads to further hydrodynamic
interactions with the boundary, with the mobility
\begin{equation}
\mu^{TR}\approx-\Lambda_{TR}^{f}\hat{h}^{-2},\qquad{\rm with}\qquad\Lambda_{TR}^{f}=\frac{1}{32\pi\eta b^{2}}\frac{1}{1+\lambda^{f}},
\end{equation}
and the propulsion coefficients of the symmetric dipole,
\begin{equation}
\pi_{1}^{(T,2s)}\approx\Lambda_{1}^{f}\hat{h}^{-2},\qquad\pi_{2}^{(T,2s)}\approx-\Lambda_{2}^{f}\hat{h}^{-2},\qquad{\rm with}\qquad\Lambda_{1}^{f}=\frac{5\lambda^{f}}{16\left(1+\lambda^{f}\right)},\quad\Lambda_{2}^{f}=\frac{5\left(2+3\lambda^{f}\right)}{48\left(1+\lambda^{f}\right)}.
\end{equation}
At this order the mean trajectory starts to be affected by the thermal
fluctuations via the convective term
\[
k_{B}T\partial_{z}\mu_{\perp}^{TT}\approx\hat{\Lambda}\hat{h}^{-2},\qquad{\rm with}\qquad\hat{\Lambda}=\frac{k_{B}T}{32\pi\eta b^{2}}\frac{2+3\lambda^{f}}{(1+\lambda^{f})}.
\]
Chemically, the effect of the flux monopole $J_{0}$ becomes apparent
with
\begin{equation}
\mathcal{E}^{(1,0)}\approx-\mathcal{E}_{1}\Lambda_{1}^{c}\hat{h}^{-2},\qquad{\rm with}\qquad\Lambda_{1}^{c}=\frac{1-\lambda^{c}}{4(1+\lambda^{c})}.
\end{equation}
This leads to the mean dynamics

\begin{align}
\left\langle \dot{x}\right\rangle  & =-\Lambda_{TR}^{f}\kappa\cos\vartheta\,\hat{h}^{-2}-\tfrac{1}{6\pi b^{3}}\mathcal{E}_{1}M_{0}J_{1}\cos\vartheta+\tfrac{9}{20\pi b^{3}}\mathcal{E}_{1}M_{1}J_{1}\sin(2\vartheta+\psi)\Lambda_{1}^{f}\hat{h}^{-2},\nonumber \\
\langle\dot{h}\rangle & =-\mu_{T}\left(1-\Lambda_{T}^{f}\hat{h}^{-1}\right)mg+\hat{\Lambda}\hat{h}^{-2}-\tfrac{1}{6\pi b^{3}}\mathcal{E}_{1}M_{0}\left(J_{1}\sin\vartheta-\Lambda_{1}^{c}J_{0}\hat{h}^{-2}\right)-\tfrac{9}{40\pi b^{3}}\mathcal{E}_{1}M_{1}J_{1}\left(3\cos(2\vartheta+\psi)-\cos\psi\right)\Lambda_{2}^{f}\hat{h}^{-2},\nonumber \\
\langle\dot{\vartheta}\rangle & =-\mu_{R}\kappa\cos\vartheta+\tfrac{3}{8\pi b^{4}}M_{1}\left(J_{1}\sin\psi+\cos\left(\vartheta+\psi\right)\Lambda_{1}^{c}J_{0}\hat{h}^{-2}\right).
\end{align}
Finally, at this order in the approximation both the parallel motion
and the orientation couple to the interface. It is evident that, at
this order, fore-aft symmetry breaking of the chemical properties of
the particle is no longer necessary for self-propulsion near a boundary.

%\bibliography{references}

\end{document}